\documentclass{aa}
\usepackage{graphics,epsfig,psfig,natbib,rotcapt,txfonts}

\def\asca{{\it ASCA}}

\def\exosat{{\it EXOSAT}} 
\def\sax{{\it BeppoSAX}} 
\def\ginga{{\it Ginga}} 
\def\rosat{{\it ROSAT}}

\def\flux{erg cm$^{-2}$ s$^{-1}$}
\def\nh{cm$^{-2}$}

\def\arcmin{$^{\prime}$}

\def\ltsima{$\; \buildrel < \over \sim \;$}
\def\simlt{\lower.5ex\hbox{\ltsima}} 
\def\gtsima{$\; \buildrel > \over \sim \;$}
\def\simgt{\lower.5ex\hbox{\gtsima}} 

\begin{document}

\title{Six years of \sax\ observations of blazars: a spectral catalog}

\author{D. Donato\inst{1} \and 
R. M. Sambruna\inst{1,2} \and M. Gliozzi\inst{1,2}}

\institute{George Mason University, School of Computational Sciences, 
4400 University Drive, Fairfax, VA 22030
\and George Mason University, Dept. Of Physics \& Astronomy, MS 3F3,
4400 University Drive, Fairfax, VA 22030}


\date{Received / Accepted 13/12/2004}

\abstract{We present a spectral catalog for blazars based on the \sax\
archive. The sample includes 44 High-energy peaked BL Lacs (HBLs), 14
Low-energy peaked BL Lacs (LBLs), and 28 Flat Spectrum Radio Quasars
(FSRQs). A total of 168 LECS, MECS, and PDS spectra were analyzed,
corresponding to observations taken in the period 1996--2002. The
0.1--50 keV continuum of LBLs and FSRQs is generally fitted 
by a single power law with Galactic column density. A minority
of the observations of LBLs (25\%) and FSRQs (15\%) is best
fitted by more complex models like the broken power law or
the continuously curved parabola. These latter models provide
also the best description for half of the HBL spectra. 
Complex models are more frequently required for sources
with fluxes $F_{\rm 2-10 keV} > 10^{-11}$\flux, corresponding
to spectra with higher signal-to-noise ratio. As a result,
considering sources with flux above this threshold,
the percentage of spectra requiring those models increases
for all the classes. We note that there is a net separation of X-ray spectral
properties between HBLs on one side, and LBLs and FSRQs on the other,
the distinction between LBLs and FSRQs is more blurry. This is most
likely related to ambiguities in the optical classification of the two
classes. 
}

\titlerunning{A \sax\ spectral catalog of blazars}
\authorrunning{D. Donato et al.\ }

\maketitle

\section{Introduction}

Blazars are a class of radio-loud Active Galactic Nuclei (AGN) defined
by non-thermal emission from a jet oriented close to the line of
sight. Multiwavelength blazar studies have made it clear that their
spectral energy distributions (SEDs) are characterized, in a $\nu$--$\nu
F(\nu)$ plot, by two broad peaks (Giommi \& Padovani 1994). The first
component, peaking anywhere in the IR--soft X--ray band, is due to
synchrotron emission, while the higher-energy one is due to the
inverse Compton emission of ambient photons off the same electrons
producing the synchrotron part of the spectrum (Maraschi et al. 1992;
Sikora et al. 1994; but see Mannheim 1993 for a different interpretation). 
 
Previous studies have shown that low-luminosity BL Lacs (High-energy peaked
BL Lacs, or HBLs) exhibit the synchrotron peak in the UV--soft
X--ray band, and the inverse Compton peak between the GeV and the TeV
band (Sambruna et al. 1994, Giommi, Ansari \& Micol 1995,
Padovani \& Giommi 1995, Fossati et al. 1997). 
The two components have approximately the same power.  
For mid-luminosity sources (Low-energy peaked
BL Lacs, or LBLs) the synchrotron peak is in the near infrared band
and the X--ray emission is due either to the synchrotron or the
Compton component, or both.  For the high-luminosity sources (Flat
Spectrum Radio Quasars, or FSRQs) the synchrotron peak is in the far
infrared band while X--ray emission is ascribed to the Compton
component.

The X-ray band, at the overlap of the synchrotron and Compton components, is
key to disentangle the contribution of the two components to the broad 
band continuum. Based on current unification studies 
(e.g., Fossati et al. 1997) one expects the low-luminosity
sources to be dominated by the high-energy tail of the synchrotron
emission in the X-rays, and thus to have steep (photon index
$\Gamma>2$, where F$_{\nu} \propto \nu^{-(\Gamma-1)}$) or convex X-ray
spectra, as a result of radiative losses. On the other hands, FSRQs
should have flatter spectra ($\Gamma<2$), while LBLs should exhibit
intermediate slopes, or even concave X-ray continua. 

Previous systematic analysis of the X-ray continua of blazars relied
on satellites with limited sensitivity (e.g., \exosat; Sambruna et
al. 1994a,b) or limited bandpass (\rosat, Urry et al. 1996, Perlman et
al. 1996; \asca, Sambruna et al. 1999, Donato et al. 2001). These
studies showed that the X-ray spectra of blazars are often complex,
with downward-curved continua in HBLs and occasionally upward-curved
continua in FSRQs and LBLs.

With its wide X--ray band pass (0.1--200 keV) and good sensitivity,
\sax\ is the instrument of choice to study the broadband X-ray
continua of blazars. The \sax\ mission ended
in early 2002; during its lifetime, \sax\ observed a large number of
blazars as part of multiwavelength campaigns or to obtain snapshot
X-ray spectra. The \sax\ observations for individual sources were
published by the original PIs of the investigation. Here we present a
uniform analysis of all the public LECS, MECS, and PDS observations,
focusing on the spectral properties. A similar previous account was
given by Giommi et al. (2002).

The paper is organized as follows. In Sect.2 we describe the sample
selection, and in Sect.3 the data reduction. In Sect.4 the spectral
catalog is described, while in Sect.5 we discuss the
results. Throughout this paper, $H_0=75$ km s$^{-1}$ Mpc$^{-1}$ and
$q_0=0.5$ are adopted.

\section{Sample Selection}

The \sax\ satellite carried onboard four Narrow Field Instruments
(NFI) pointing in the same direction and covering a very large energy
range from 0.1 to 300 keV (Boella et al. 1997). Two of the four
instruments have imaging capability, the Low Energy Concentrator
Spectrometer (LECS), sensitive in the nominal range 0.1--10 keV, and three
Medium Energy Concentrator Spectrometers (MECS), sensitive in the
nominal range 1--10 keV.  The other two detectors are the High Pressure
Proportional Counter (HPGSPC), sensitive in the nominal range 4--120 keV, and
the Phoswich Detector System (PDS), sensitive in the nominal range 13--300
keV. We restricted our analysis to the LECS, MECS, and PDS archives,
as the HPGSPC has not enough sensitivity to detect relatively faint
sources as blazars.

We cross-correlated the \sax\ database with published lists of
blazars, including Fossati et al. \cite{fossati}, Donato et al. 
\cite{donato}, and Giommi et al. \cite{giommi}, and references
herein. Following Padovani \& Giommi \cite{padogiommi}, sources with
the radio-to-X-ray spectral index, $\alpha_{rx}$, smaller than 0.75
were classified as HBLs, while sources with $\alpha_{rx}$\gtsima0.75
were classified as LBLs or FSRQs, depending on the EW of the
optical emission lines.

The sample is presented in Table~\ref{tab:catalog}, 
where we list the name of the
object (Col. 1), the coordinates for the equinox 2000.0 (Cols. 2 and
3), the redshift (Col. 4), the Galactic column density $N_{\rm H}^{\rm
Gal}$ (Col. 5), and the number of \sax\ observations of the source
(Col. 6). The Galactic column was derived from Dickey \& Lockman
(1990). The sample includes 86 blazars: 44 HBLs, 14 LBLs, and 28
FSRQs. Most sources were observed as part of multiwavelength
campaigns, or in a single snapshot observation to obtain the Spectral
Energy Distributions. Since the \sax\ selection process favored
objects known for their high X--ray brightness, the sample is highly
heterogeneous (with some sources observed multiple times), strongly
biased, and by no means complete.

\section{Observations and Data Analysis}

All public observations up to January 2002 were analyzed, for a total
of 168 spectra. In Table~\ref{tab:log} we present the log of the observations,
together with the exposure times and the mean count rates in the
various instruments.  For each observation we report the LECS count
rates in the energy range 0.1--2 keV and the MECS count rates in the
energy range 2--10 keV.
Inspection of the PDS spectra shows that the background usually
dominates above 50 keV. Thus, counts were extracted in the energy range 
13--50 keV for all sources. However, for the brightest sources reliable
counts up to 200 keV were detected. In these cases, spectral fits to the PDS 
data were performed in the energy range 0.1--200 keV. 

In Table~\ref{tab:log}, 
we report the PDS count rates for only those sources which
were detected at \gtsima 3$\sigma$ confidence level. 
Most of the sources are weak at energies greater than 10 keV and
only 36 sources are detected (15 HBLs, 4 LBLs, and 17 FSRQs). 

We extracted LECS, MECS, and PDS spectra. 
The data analysis for the LECS and MECS instruments was based on
the linearized, cleaned event files obtained from the on-line archive.
The spectra were extracted with the FTOOLS package XSELECT (v. 2.2),
using extraction regions of radius of 8\arcmin\ and 4\arcmin\ for the
LECS and MECS, respectively.  In the case of weak sources, we used a
radius of 6\arcmin\ and sometimes even 4\arcmin\ to extract the LECS
spectra. 
The LECS and MECS background is low but not
uniformly distributed across the detector. For this reason it is
better to evaluate the background from blank fields that are available
from the SDC public ftp site.  Since these background data were taken
in regions of the sky with no detected sources and low Galactic
absorption, the low energy X--ray counts coming from sources located
in regions with high Galactic absorption may be underestimated and/or
the intrinsic absorption may be overestimated.

The spectral analysis was performed with XSPEC v.11.2, using the
latest available response matrices from the \sax\ calibration center.
The spectra from LECS, MECS, and PDS detectors were rebinned using the
channel grouping suggested by the \sax\ team which was designed to
match the detector resolution and sensitivity in an optimal way.  The
LECS/MECS and MECS/PDS normalization factors (to account for
inter-calibration systematics of the instruments) were left free to
vary; normal acceptable values are in the range 0.65--1.0 and
0.77--0.93, respectively (Fiore et al. 1999). Spectral fits were
performed in the energy ranges 0.1--2 keV for the LECS, 2--10 keV for
the MECS, and 13--200 keV for the PDS where the calibration is best
known and the background contribution negligible.  The best-fit models
were determined using the $\chi^2$ minimization routine. The
significance of the fit improvement, when additional free parameters
were added, was evaluated using the F-test, assuming as a threshold
for significant improvement P$_F$=95\%. Uncertainties on the fitted
parameters are 90\% confidence ($\Delta\chi^2$=2.7) for one parameter
of interest. 

The following spectral models were used to fit the \sax\ spectra: 

\noindent (1) A single power law with photon index $\Gamma$ and column
density N$_H$ fixed to the Galactic value; 

\noindent (2) A single power law with free N$_H$; 

\noindent (3) A broken power law with break energy E$_{\rm break}$, and
photon indices below and above the break $\Gamma_1$ and $\Gamma_2$,
respectively; 

\noindent (4) A continuously curved parabola (Fossati et
al. 2000). The latter has the same parameters as model 3), but
$\Gamma_1$ and $\Gamma_2$ are evaluated at 1 keV and 10 keV,
respectively. 

We used the following procedure. At first, all the spectra were fitted
with model (1). When the fit was unsatisfactory, model (2) was
used. In several cases, the fitted column density from model (2) was
consistent with the Galactic value within the errors. We interpret
this result as marginal evidence for continuum curvature; however, in
these cases fits with the curved models (3) and (4) did not provide a
significant improvement. The observations for which models (1) and (2)
provided the best-fit are reported in Table~\ref{tab:bestfit}. 

However, in the case of bright sources, $F_{\rm 2-10 keV} > 10^{-11}$\flux,
corresponding to spectra with higher signal-to-noise ratio, a curved
model was generally found to provide a better description of the \sax\
spectra than either model (1) or (2). 
In few cases (6 observations) where the curved model with Galactic absorption
did not provide a good description of the spectrum, the absorption was left 
free to vary. These observations are listed in Table~\ref{tab:bestfit2}. 
In Table~\ref{tab:powerlaw} we also list the fits with models (1) and (2) 
for the same spectra, for future use. As a summary, the last column of 
Table~\ref{tab:log} 
reports the model that best-fits the individual spectra for each
source in the sample. 

The choice of the above spectral models is motivated by previous X-ray
studies of blazars (e.g., \exosat, \ginga, \asca; Sambruna et
al. 1994a; Tashiro et al. 1995; Takahashi et al. 1996; Donato et
al. 2001). These authors found that the X-ray continua of
high-luminosity blazars are usually well described by a single power
law, while at decreasing luminosities curved X-ray continua are often
observed. The latter are convex, with $\Gamma_1 < \Gamma_2$, in HBLs
and concave, with $\Gamma_1 > \Gamma_2$, in some LBLs and FSRQs.  In
the brightest HBLs, the broken power law model is inadequate to
represent the continuous curvature of the X-ray spectra, and better
results are obtained using a continuously curved model (i.e., Fossati
et al. 2000 for MKN~421, Tavecchio et al. 2001 for MKN~501).

An independent analysis of the \sax\ blazar archive was performed by
Giommi et al. (2002), who analyzed a total of 157 X-ray spectra of 84
sources (42 HBLs, 12 LBLs, 22 FSRQs, and 4 GigaHertz Peaked Spectrum
QSOs). These authors fitted the LECS+MECS+PDS spectra using a single power
law, a broken power law, a sum of two power laws, and a logarithmic
parabola. For each model the $N_{\rm H}$ was fixed at the Galactic
value. Comparing the results for the models that are in common with
this work (single power law or broken power law), we find that our
results are completely consistent with Giommi et al. (2002). 

\section{The Spectral Catalog}

The results of our spectral analysis, from fits to the joint
LECS+MECS+PDS datasets in the energy range 0.1--200 keV, are reported
in Table~\ref{tab:bestfit} and \ref{tab:bestfit2}. For each source we present
the parameters obtained with the model that best-fits the data
(power-law model in Table~\ref{tab:bestfit} and curved model in 
Table~\ref{tab:bestfit2}), and
their 90\% confidence uncertainties. For sources with multiple
observations, different models best-fit the data at different
epochs. We reported the individual best-fit models, using a
different procedure than Giommi et al. \cite{giommi}, who
systematically used the same model for repeatedly observed sources.

For the HBL class we obtained spectral information for 90 observations
of 44 sources. For 46 observations (corresponding to 15 sources) the
model that best-fits the LECS+MECS+PDS data is a downward-curved model
(broken power law or curved parabola), while the remaining  
observations are described by a single power law. In the latter cases, the
photon index is steep, $\Gamma$\gtsima 2.  The energy break, in the
cases of a curved model fit, is more frequently located around 1-2
keV. Although for most of the sources the best fit is obtained with
intrinsic absorption fixed to the Galactic value, 20 spectral fits
corresponding to 17 sources require a column density of the order 
of $10^{20}$\nh. This may indicate residual curvature in the continuum
which the model is still inadequate to describe. 

For the LBL class we analyzed 25 observations of 14 sources. For 7
observations (corresponding to 4 sources) the model that best-fits the
LECS+MECS+PDS data is a curved model, while the remaining observations
are described by a single power law. The photon indices of the single
power laws are flat, $\Gamma < 2$, in 9 sources (13 observations) and
steep, $\Gamma \geq 2$, in 4 sources (4 observations).  For the
sources with a spectrum described by a curved model, S5~0716+71 and
ON~231 need for all their observations an upward curvature, with a
steep photon index below the break ($\Gamma_1 \geq 2.4$) and a flatter
index above the break ($\Gamma_2 \leq 2$).  One observation of 3C~66A
and one of BL~Lac need a downward-curved model. In both sources,
$\Gamma_1 \sim$2.2, while $\Gamma_2 \sim$ 2.3 and 2.6, respectively,
while the break energy is at 0.3--0.4 keV.  Among the sources with
multiple observations, only BL~Lac shows spectral variability, with a
total change of the photon index $\Delta\Gamma \sim 1$ for a change of
the flux of a factor 3 (from 6 to 20$\times 10^{-12}$ \flux).

Finally, for the FSRQ class 53 observations (corresponding to 28
sources) were analyzed. The best-fit is always obtained using a single
power law (model (1)) with $\Gamma \leq 1.8$, except for 1ES~0836+710,
PKS~1510-089, PKS~2126-158, and 5 observations of 3C~273, for which a
curved model is needed. All the observations are fitted using the
intrinsic absorption fixed to the Galactic value.  For 1ES~0836+710
and PKS~2126-158 downward curvature is detected, with $\Gamma_1 \sim
1$ and $\Gamma_2 \sim 1.3$ and 1.8, respectively.  For 3C~273 and
PKS~1510-089 upward curvature is indicated, indicating the presence of
a soft excess. 

Since the soft excess can have thermal origins (e.g., accretion disk),
we fitted the \sax\ data with a power law plus either a blackbody or a
bremsstrahlung. The inclusion of either thermal component was not
statistically preferred to a broken power law. Thus, the origin of 
the soft excess in 3C~273 and PKS~1510-089 is still an open question. 

\section{Results and Discussion} 

\begin{table}
{\bf Table 6.} Average spectral parameters for the best-fit models 
\label{tab:average}
\begin{center}
\begin{tabular}{l|ccc}
\hline
\hline
&HBL&LBL&FSRQ\\
\hline
&\multicolumn{3}{c}{a) Single power law}\\
N                      &      34       &    12       &    26      \\
$<\Gamma_{\rm PL}>$    &   2.24(0.04)  &  1.92(0.07) &  1.59(0.05)\\
\hline
&\multicolumn{3}{c}{b) Broken power law}\\
N                      &       6       &     2       &     2      \\
$<\Gamma_1>$           &   2.06(0.18)  &  2.29(0.07) &  1.06(0.03)\\
$<E_{\rm break}>$      &   2.24(1.02)  &  1.10(0.65) &  0.49(0.11)\\
$<\Gamma_2>$           &   2.65(0.26)  &  2.24(0.40) &  1.56(0.22)\\
\hline
&\multicolumn{3}{c}{c) Continuously curved parabola}\\
N                      &       7       &     1       &     1      \\
$<\Gamma_1>$           &   1.91(0.14)  &  2.58       &  2.51      \\
$<E_{\rm break}>$      &   1.71(0.36)  &  3.10       &  1.40      \\
$<\Gamma_2>$           &   2.53(0.12)  &  1.58       &  1.32      \\
\hline
\end{tabular}

\end{center}
The break energy is in keV. For each value we report also the standard
deviation (in parenthesis) and the number N of observations used. 
  
\end{table}

\begin{table}
{\bf Table 7.} Average parameters for the blazar classes 
\label{tab:average2}
\begin{center}
\begin{tabular}{l|ccc}
\hline
\hline
&HBL&LBL&FSRQ\\
\hline
N                &      47       &    15       &    29      \\
\hline
$<\Gamma> $      &   2.27(0.04)  &  2.01(0.08) &  1.58(0.05)\\
\hline
$<L_{\rm X}>$    &  44.61(0.11)  & 44.49(0.27) & 45.85(0.22)\\
\hline
\end{tabular}

\end{center}
Average values of the X--ray spectral indices and monochromatic 
1 keV luminosities from fitting all the \sax\ spectra with either
model (1) or model (2). The standard deviation are in parenthesis.
  
\end{table}

We presented a uniform analysis of the \sax\ archival spectra for
blazars. We will now use the spectral catalog to infer the average
X-ray properties of the three blazar classes. 
While these properties were investigated
before with \asca\ (Sambruna et al. 1999, Donato et al. 2001), only
\sax\ has the unique combination of wide bandpass and sensitivity
necessary to study the entire X-ray range. 
An important caveat, however, is that
the ``sample'' is incomplete and by all means biased toward the
brightest X-ray sources of each class. 

First, we derive average X-ray spectral parameters for each best-fit
model and each class of blazars. To avoid a bias toward sources
observed multiple times, we adopted the following procedure. For
sources observed only a few times ($\leq 9$) and with negligible
variations in spectral index ($\Delta \Gamma$ \ltsima 0.5) or flux
variation of a factor of $\sim2$ between maximum and minimum, we used
the observation for which the spectral parameters are better
constrained.
For sources observed multiple
times and with significant spectral or flux variations, we considered
two observations, corresponding to the two most extreme values of the
spectral index and/or flux.  This procedure yielded 47 observations of
44 HBLs (Mkn 421, Mkn 501, and 1ES2344+514 were observed in 2 states),
15 observations of 14 LBLs (BL Lac showed variability both in the flux
and in the spectral index), and 29 observations of the 28 FSRQs
(PKS~0528+134 showed variability in the spectral index). The average
parameters are reported in Table~\ref{tab:average}.

\begin{figure}[h]
\begin{center}
\noindent{\psfig{file=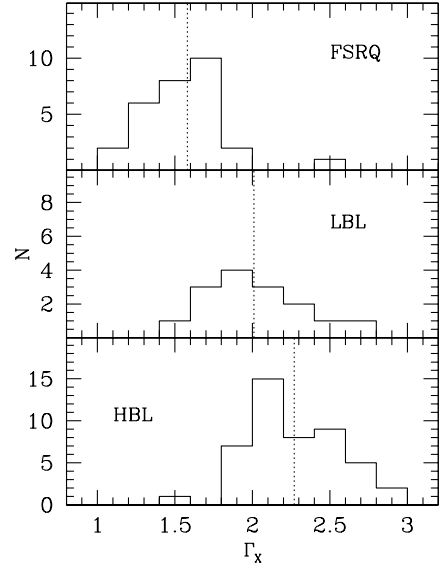,height=7.5cm,bbllx=55pt,bblly=157pt,bburx=460pt,bbury=690pt,clip=}}
\caption[]{Distribution of the photon spectral indices obtained
using a single power law fit.
The dotted line represent the average value reported in 
Table~\ref{tab:average2}.}  
\label{fig:istogamma}
\end{center}
\end{figure}

\begin{figure}[h]
\begin{center}
\noindent{\psfig{file=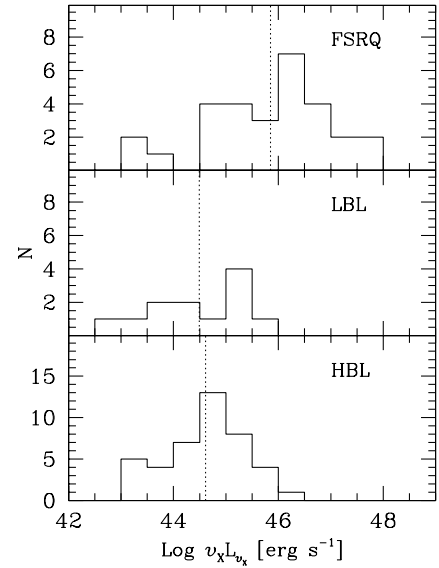,height=7.5cm,bbllx=55pt,bblly=150pt,bburx=460pt,bbury=690pt,clip=}}
\caption[]{Distribution of the $\nu L\nu$ monochromatic (1 keV) luminosity 
for the three subclasses obtained
using a single power law fit.
The dotted line is the average spectral index from Table~\ref{tab:average2}.} 
\label{fig:istolum}
\end{center}
\end{figure}

Second, in order to compare in a homogeneous way all the objects, we
parameterize the continuum in 0.1--200 keV with a simple power law
plus {\it fixed} and {\it free} absorption, even for sources whose
spectrum was better described by a curved model.  Thus, we used the
values of the photon index and luminosity from the fits reported in
Table~\ref{tab:bestfit} and ~\ref{tab:powerlaw}. 
Using this approach, a convex/concave continuum will
be represented by a power law with larger/smaller absorption column
than Galactic. To reduce the bias toward sources with multiple
observations, we adopted the procedure described above which reduces
the number of observations to a maximum of 2 for the most variable
sources.

\begin{figure}[th]
\begin{center}
\noindent{\psfig{file=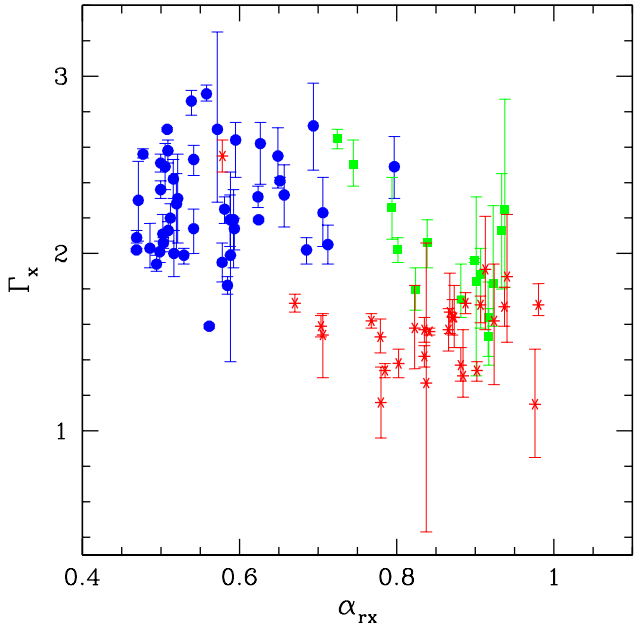,width=8.5cm,bbllx=60pt,bblly=190pt,bburx=550pt,bbury=680pt,clip=}}
\caption[]{X-ray spectral indices vs. broad band radio to X--ray indexes. Circle: HBLs, Square: LBLs, Star: FSRQs.}
\label{fig:gammaalfarx}
\end{center}
\end{figure}

\begin{figure}[th]
\begin{center}
\noindent{\psfig{file=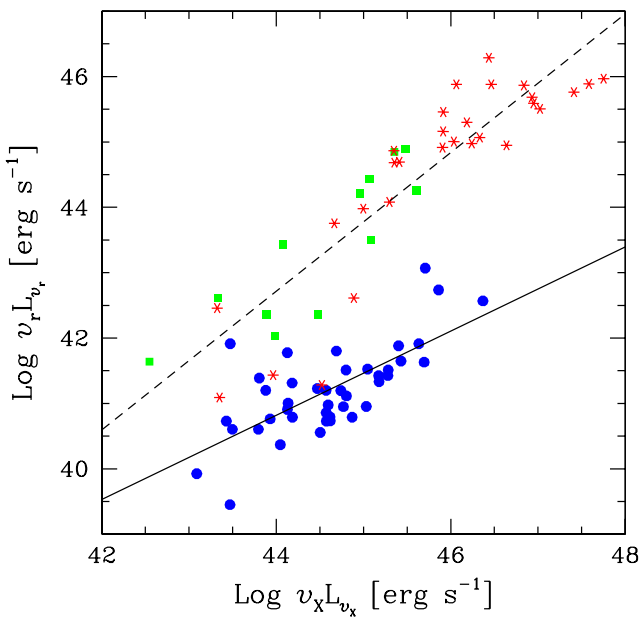,width=8.5cm,bbllx=45pt,bblly=180pt,bburx=560pt,bbury=680pt,clip=}}
\caption[]{Monochromatic X-ray (1 keV) luminosities vs. monochromatic radio (5 GHz) luminosities. Circle: HBLs, Square: LBLs, Star: FSRQs.
Linear correlations are plotted for the HBL class (continuous line) and
for the LBL-FSRQ classes (dashed line).}
\label{fig:lumxlumr}
\end{center}
\end{figure}

The distributions of the spectral indices from the fits with a single
power law model are shown in Figure~\ref{fig:istogamma} for 
the three classes. Figure~\ref{fig:istolum}
shows the distributions for the monochromatic luminosities at 1
keV. The average values are reported in Table~\ref{tab:average2}.  
To compare the distributions of the 3 classes in Figure~\ref{fig:istogamma} 
and ~\ref{fig:istolum}, we used a
Kolmogorov-Smirnov test, which gives the probability P$_{KS}$ that 2
distributions are drawn from the same parent population.  Therefore,
small values of P$_{KS}$ indicate that the cumulative distribution
function of the first data set is significantly different from that of
the second data set.  For the spectral index distributions in 
Figure~\ref{fig:istogamma}, P$_{KS} \sim$ 0.23 for HBLs vs. LBLs, 
P$_{KS} \sim$ 0.99 for HBLs
vs. FSRQs, and P$_{KS} \sim$ 0.83 for LBLs vs. FSRQs. Thus, we
conclude that there is no significant difference in the spectral index
distribution among the three classes. However, the Kolmogorov-Smirnov
test is sensitive only to the shape of the distribution, and not to
the location of its centroid.  We thus 
compared the average values of the spectral index for the 3 subclasses in
Table~\ref{tab:average2}. Comparing the average indices for HBLs and LBLs and
for LBLs and FSRQs we found that they are marginally consistent at 
2$\sigma$ and 3$\sigma$, respectively. On the other hand, 
the average spectral index of HBLs is significantly different from 
the FSRQs, with HBLs having a steeper X-ray continua than FSRQs.

Similar conclusions are derived if the average luminosities of the
three classes are compared. Using the Kolmogorov-Smirnov test, we find
P$_{KS} \sim$ 0.31 for HBLs vs. LBLs, P$_{KS} \sim$ 0.89 for HBLs
vs. FSRQs, and P$_{KS} \sim$ 0.31 for LBLs vs. FSRQs, suggesting that
the distribution shapes are drawn from the same population. Comparing
the average luminosities shows that HBLs and FSRQs are different at
3$\sigma$, while HBLs and LBLs have similar $\langle L_{1~keV}
\rangle$. 

Using published radio fluxes at 5 GHz, we derived the radio-to-X-ray
indices $\alpha_{\rm rx}$. In Figure~\ref{fig:gammaalfarx} 
we plot the X-ray photon index from Table~\ref{tab:bestfit2} 
versus $\alpha_{\rm rx}$. The blazars separate in two
groups: HBLs have $\alpha_{\rm rx} < 0.75$ and $\Gamma_x > 2$, while
LBLs and FSRQs have $\alpha_{\rm rx} > 0.75$ and $\Gamma_x < 2$. One
LBL (BL Lac) and 4 FSRQs (WGA J0546.6-6415, RGBJ1629+4008,
RGBJ1722+2436, and S5 2116+81) fill the gap between the two groups. As
discussed in Padovani et al. \cite{padovani}, these sources exhibit
spectral energy distributions more similar to HBLs and were dubbed
``High-energy peaked FSRQs''. It still remains to be demonstrated
whether the steep soft X-ray spectrum of these sources is due to
synchrotron from the jet or to is thermal emission perhaps related to the
high-energy tail of the disk Blue Bump (as, e.g., in 3C~273).
 
Figure~\ref{fig:lumxlumr} shows the plot of the radio luminosity versus 
the X-ray luminosity for HBLs, LBLs, and FSRQs. There is a clear correlation 
between the luminosities at the
two wavelengths. However, different trends are observed between HBLs
and LBLs-FSRQs: for a given X-ray luminosity, LBLs and FSRQs appear
more luminous at radio. 

To quantify the degree of linear correlation, we calculated the linear
correlation coefficient $r$ and computed the chance probability
$P_{\rm r}(N)$ that a random sample of $N$ uncorrelated pairs of
measurements would yield a linear correlation coefficient equal or
larger than $|r|$; if the chance probability is small, the two
quantities are likely to be correlated.  For HBLs, the linear
correlation coefficient is $r=0.72$ and the chance probability $P_{\rm
r}=2 \times 10^{-7}$. For LBLs-FSRQs, $r=0.90$ and $P=2.8 \times
10^{-15}$.  From a linear fit to the data in Figure~\ref{fig:lumxlumr}
(i.e., considering
the logarithmic values for the luminosities), we derive the following
expressions: $Log L_{\rm X}$=$12.5+0.64*Log L_{\rm R}$ for HBLs, and
$Log L_{\rm X}$=$-3.9+1.06*Log L_{\rm R}$ for LBLs-FSRQs.   These results
should be taken with caution, since our sample is very likely biased
toward brighter X-ray sources. 

In closing, we comment on flux and spectral variability. More details
can be found in the publications for individual sources from the
original PIs of the observations. Several sources in our sample were
observed repeatedly with \sax. Inspection of the plots of flux versus
spectral index shows that significant flux and spectral variability is
present only for the HBL class, with the familiar trend of flatter
slope with increasing flux. This confirms previous results based on
various X-ray satellites (e.g., Mkn~421 with \asca, Takahashi et
al. 1999).

In summary, we presented an analysis of the \sax\ database for
blazars. 
Most HBLs have downward-curved continua, in agreement
with the synchrotron interpretation, while the brightest FSRQs and
LBLs show upward curvatures. 
While there is a net separation between HBLs on one side, and LBLs 
and FSRQs on the other, the distinction between LBLs and FSRQs is more
blurry. Conceivably, this is due to optical identification
ambiguities, as the classification of a higher-luminosity blazar as
an LBL or FSRQ depends on the Equivalent Width of the optical emission
lines; thus, variability of the optical continuum makes this classification
highly dependent on the observation epoch.

\begin{acknowledgements}
We gratefully acknowledge financial support from NASA grants
NAG5-10073 (DD, RMS) and LTSA grant NAG5-10708 (RMS, MG).  RMS was
also supported by an NSF CAREER award and the Clare Boothe Luce
Program of the Henry Luce Foundation.  This research made use of 
data retrieved from the ASI/ASDC-\sax public archive
and the NASA/IPAC Extragalactic Database (NED) which is operated by 
the Jet Propulsion Laboratory, Caltech, under contract with the 
National Aeronautics and Space Administration.
\end{acknowledgements}

\onecolumn
\begin{table*}
\caption{The sample}
\label{tab:catalog}
\begin{center}
\scriptsize
\begin{tabular}{lcrccc}
\hline
\noalign{\smallskip}
\hline
\noalign{\smallskip}
~~~~~Obj. Name     & R.A.(J2000)    &  Dec.(J2000)    & Redshift & $N_{\rm H,Gal}$ & No. of  \\
                   &                &                 &          &  $10^{20}$      & observ. \\
~~~~~~~~~~(1)      & (2)            &  (3)~~~~~~      &  (4)     & (5)             & (6)     \\
\noalign{\smallskip}
\hline
\noalign{\smallskip}
\multicolumn{6}{c}{HBLs}\\
\noalign{\smallskip}
\hline
\noalign{\smallskip}
1ES0033+595        &   00 35 52.5   &   +59 50 03.9   & 0.086 & 42.7  &    \\
1ES0120+340        &   01 23 08.5   &   +34 20 48.9   & 0.272 &  5.14 & 2  \\
RXJ0136.5+3905     &   01 36 32.8   &   +39 05 56.0   & --    &  6.01 &    \\
1ES0145+138        &   01 48 29.8   &   +14 02 16.0   & 0.125 &  5.18 &    \\
MS0158.5+0019      &   02 01 06.2   &   +00 34 00.0   & 0.298 &  2.67 &    \\
1ES0229+200        &   02 32 48.4   &   +20 17 16.0   & 0.140 &  9.32 &    \\
MS0317.0+1834      &   03 19 51.8   &   +18 45 35.0   & 0.190 & 10.4  &    \\
1ES0323+022        &   03 26 13.9   &   +02 25 14.9   & 0.147 &  8.81 &    \\
1ES0347-121        &   03 49 23.2   &   -11 59 27.0   & 0.185 &  3.63 &    \\
1ES0414+009        &   04 16 52.4   &   +01 05 24.0   & 0.287 & 10.5  &    \\
1ES0502+675        &   05 07 56.2   &   +67 37 24.0   & 0.314 &  9.11 &    \\
1ES0507-040        &   05 09 38.2   &   -04 00 46.0   & 0.304 &  9.74 &    \\
PKS0548-322        &   05 50 40.6   &   -32 16 14.9   & 0.069 &  2.20 & 3  \\
MS0737.9+7441      &   07 44 05.2   &   +74 33 58.0   & 0.315 &  3.29 &    \\
B20912+29          &   09 15 52.3   &   +29 33 20.0   & --    &  1.97 &    \\
1ES0927+500        &   09 30 37.6   &   +49 50 26.9   & 0.188 &  1.40 &    \\
1ES1028+511        &   10 31 18.4   &   +50 53 36.0   & 0.361 &  1.15 &    \\
RXJ1037.7+5711     &   10 37 44.2   &   +57 11 54.9   & --    &  0.55 &    \\
RXJ1058.6+5628     &   10 58 37.6   &   +56 28 11.9   & 0.144 &  0.68 & 2  \\
1ES1101-232        &   11 03 37.6   &   -23 29 30.9   & 0.186 &  6.05 & 2  \\
Mkn421             &   11 04 27.3   &   +38 12 32.0   & 0.030 &  1.34 & 16 \\
RXJ1117.1+2014     &   11 17 06.1   &   +20 14 07.0   & 0.139 &  1.33 &    \\
1ES1118+424        &   11 20 48.0   &   +42 12 12.0   & 0.124 &  1.93 &    \\
1ES1133+704        &   11 36 26.4   &   +70 09 28.0   & 0.045 &  1.40 &    \\
RXJ1211.9+2242     &   12 11 58.0   &   +22 42 36.0   & 0.455 &  2.22 & 3  \\
ON325              &   12 17 52.0   &   +30 07 00.9   & 0.130 &  1.70 & 2  \\
1ES1218+304        &   12 21 21.7   &   +30 10 36.9   & 0.182 &  1.71 &    \\
1ES1255+244        &   12 57 31.9   &   +24 12 38.9   & 0.141 &  1.34 &    \\
MS1312.1-4221      &   13 15 03.7   &   -42 36 50.0   & 0.108 &  8.45 &    \\
1RXSJ141756.8+25   &   14 17 56.7   &   +25 43 28.9   & 0.237 &  1.58 & 3  \\
1ES1426+428        &   14 28 32.5   &   +42 40 18.9   & 0.129 &  1.36 &    \\
MS14588+2249       &   15 01 02.8   &   +22 37 54.9   & 0.235 &  3.54 &    \\
1ES1517+656        &   15 17 47.6   &   +65 25 23.9   & 0.702 &  1.93 &    \\
1ES1533+535        &   15 35 00.7   &   +53 20 38.0   & 0.890 &  1.32 &    \\
1ES1544+820        &   15 40 15.5   &   +81 55 04.0   & --    &  4.32 &    \\
1ES1553+113        &   15 55 43.1   &   +11 11 20.0   & 0.360 &  3.67 &    \\
Mkn501             &   16 53 52.2   &   +39 45 37.0   & 0.034 &  1.81 & 11 \\
H1722+119          &   17 25 04.3   &   +11 52 15.9   & 0.018 &  8.86 &    \\
1ES1741+196        &   17 43 57.7   &   +19 35 08.9   & 0.084 &  6.68 &    \\
1ES1959+650        &   19 59 59.8   &   +65 08 54.9   & 0.047 &  9.89 & 3  \\
PKS2005-489        &   20 09 25.3   &   -48 49 53.0   & 0.071 &  5.09 & 2  \\
PKS2155-304        &   21 58 51.7   &   -30 13 28.9   & 0.116 &  1.70 & 3  \\
1ES2344+514        &   23 47 04.8   &   +51 42 17.9   & 0.044 & 15.4  & 7  \\
H2356-309          &   23 59 07.6   &   -30 37 36.9   & 0.165 &  1.37 &    \\
\hline
\noalign{\smallskip}
\multicolumn{6}{c}{LBLs}\\
\noalign{\smallskip}
\hline
\noalign{\smallskip}
PKS0048-09         &   00 50 41.2   &   -09 29 05.9   & --    &  3.85 &    \\
3C66A              &   02 22 39.6   &   +43 02 08.0   & 0.444 &  9.41 & 2  \\
AO0235+164         &   02 38 38.8   &   +16 36 59.0   & 0.940 &  8.85 &    \\
PKS0537-441        &   05 38 50.3   &   -44 05 08.9   & 0.894 &  3.94 &    \\
S50716+71          &   07 21 53.4   &   +71 20 35.9   & --    &  3.84 & 3  \\
OJ287              &   08 54 49.0   &   +20 06 32.0   & 0.306 &  3.05 & 2  \\
PKS1144-379        &   11 47 01.4   &   -38 12 11.0   & 1.048 &  7.53 &    \\
ON231              &   12 21 31.6   &   +28 13 58.0   & 0.102 &  1.88 & 2  \\
OQ530              &   14 19 46.6   &   +54 23 16.0   & 0.151 &  1.18 & 3  \\
PKS1519-273        &   15 22 37.7   &   -27 30 11.0   & 0.071 &  8.41 &    \\
S51803+784         &   18 00 45.7   &   +78 28 04.0   & 0.680 &  3.94 &    \\
3C371              &   18 06 50.6   &   +69 49 27.9   & 0.051 &  4.74 &    \\
4C56.27            &   18 24 07.1   &   +56 51 00.0   & 0.664 &  4.14 &    \\
BLLAC              &   22 02 43.3   &   +42 16 40.0   & 0.069 & 20.1  & 5  \\
\hline
\noalign{\smallskip}
\multicolumn{6}{c}{FSRQs}\\
\noalign{\smallskip}
\hline
\noalign{\smallskip}
PKS0208-512        &   02 10 46.1   &   -51 01 01.9   & 0.999 &  3.09 &    \\
NRAO140            &   03 36 29.8   &   +32 18 20.0   & 1.258 & 15    &    \\
PKS0521-365        &   05 22 58.0   &   -36 27 30.9   & 0.055 &  3.45 &    \\
PMN0525-3343       &   05 25 06.2   &   -33 43 05.0   & 4.401 &  2.24 &    \\
PKS0528+134        &   05 30 56.4   &   +13 31 55.0   & 2.060 & 25.7  & 8  \\
WGAJ0546.6-6415    &   05 46 41.8   &   -64 15 21.9   & 0.323 &  4.44 &    \\
PKS0743-006        &   07 45 54.1   &   -00 44 17.0   & 0.994 &  6.48 &    \\
1ES0836+710        &   08 41 24.4   &   +70 53 41.9   & 2.172 &  2.78 &    \\
RGBJ0909+039       &   09 09 15.8   &   +03 54 42.1   & 3.200 &  3.48 &    \\
PKS1127-14         &   11 30 07.0   &   -14 49 27.1   & 1.187 &  3.83 &    \\
3C273              &   12 29 06.7   &   +02 03 09.0   & 0.158 &  1.79 & 9  \\
3C279              &   12 56 11.1   &   -05 47 22.0   & 0.536 &  2.26 & 5  \\
GB1428+4217        &   14 30 23.8   &   +42 04 36.9   & 4.715 &  1.38 &    \\
GB1508+5714        &   15 10 02.8   &   +57 02 44.9   & 4.301 &  1.40 & 2  \\
PKS1510-089        &   15 12 50.4   &   -09 05 59.9   & 0.360 &  8.18 &    \\
\hline
\end{tabular}
\end{center}
\end{table*}
\begin{table*}
\begin{center}
\scriptsize
\begin{tabular}{lccccc}
\hline
\noalign{\smallskip}
\hline
\noalign{\smallskip}
~~~~~Obj. Name     & R.A.(J2000)    &  Dec.(J2000)    & Redshift & $N_{\rm H,Gal}$ & No. of \\
                   &                &                 &          & $10^{20} {~\rm cm}^{-2}$       & observ. \\
~~~~~~~~~~(1)      & (2)            &  (3)            &  (4)     & (5)             & (6)     \\
\noalign{\smallskip}
\hline
\noalign{\smallskip}
RGBJ1629+4008      &   16 29 01.3   &   +40 07 58.0   & 0.272 &  0.84 &    \\
1ES1641+399        &   16 42 58.8   &   +39 48 37.0   & 0.593 &  1.01 &    \\
RGBJ1722+2436      &   17 22 41.3   &   +24 36 19.0   & 0.175 &  4.71 &    \\
S41745+62          &   17 46 13.9   &   +62 26 56.0   & 3.889 &  3.43 &    \\
S5 2116+81         &   21 14 01.2   &   +82 04 47.9   & 0.084 &  7.60 & 2  \\
PKS2126-158        &   21 29 12.1   &   -15 38 41.9   & 3.268 &  4.92 &    \\
PKS2134+004        &   21 36 38.5   &   +00 41 53.9   & 1.932 &  4.48 &    \\
PKS2149-306        &   21 51 55.3   &   -30 27 53.9   & 2.345 &  2.11 &    \\
PKS2223+210        &   22 25 37.9   &   +21 18 06.9   & 1.959 &  4.37 &    \\
PKS2223-05         &   22 25 47.2   &   -04 57 02.9   & 1.404 &  5.65 &    \\
H2230+114          &   22 32 36.4   &   +11 43 50.9   & 1.037 &  5.00 & 5  \\
PKS2243-123        &   22 46 18.1   &   -12 06 50.0   & 0.630 &  4.78 &    \\
H2251+158          &   22 53 57.7   &   +16 08 53.9   & 0.859 &  6.55 &    \\

\hline
\end{tabular}
\end{center}
{\bf Columns}: {\bf 1}=Object Name as found in \sax\ archive; 
{\bf 2}=Right Ascension (at J2000); {\bf 3}=Declination (at J2000);
{\bf 4}=Redshift; {\bf 5}=Galactic absorption based on Dickey \& Lockman \cite{dickey}; 
{\bf 6}=Number of observations in the \sax\ archive.
\end{table*}

\clearpage
\begin{table*}
\caption{Observation Log}
\label{tab:log}
\begin{center}
\scriptsize
\begin{tabular}{lccccccccc}
\hline
\noalign{\smallskip}
\hline
\noalign{\smallskip}
~~~~~Obj. Name     & Obs. Date   & Seq. Num & LECS     &    c/s     & MECS      &      c/s   &  PDS      &     c/s       & Model \\
                   &             &          & (ksec)  & ($10^{-2}$)  &  (ksec)  &  ($10^{-2}$) &  (ksec)  &  ($10^{-2}$) &       \\
~~~~~~~~~~(1)      & (2)         &  (3)     &  (4)     & (5)        &   (6)     &  (7)       & (8)       & (9)           &  (10) \\
\noalign{\smallskip}
\hline
\noalign{\smallskip}
\multicolumn{9}{c}{HBLs}\\
\noalign{\smallskip}
\hline
\noalign{\smallskip}

1ES0033+595      & 18-Dec-1999 &  50863001  &  13.5 &  13.92$\pm$  0.33 &  43.5 &  61.70$\pm$  0.38 &  22.9 &   35.0$\pm$   2.8 & 2 \\
1ES0120+340      & 03-Jan-1999 &  50493001  &   9.9 &  12.09$\pm$  0.36 &  32.1 &  19.26$\pm$  0.25 &  16.3 &   20.1$\pm$   3.5 & 2 \\
                 & 02-Feb-1999 &  504930011 &   4.9 &  11.73$\pm$  0.50 &  21.7 &  14.14$\pm$  0.26 &  10.1 &   16.8$\pm$   4.4 & 2 \\
RXJ0136.5+3905   & 09-Jan-2001 &  51241003  &  19.8 &  11.07$\pm$  0.24 &  48.1 &  12.99$\pm$  0.17 &  22.0 &          ...      & 2 \\
1ES0145+138      & 30-Dec-1997 &  50064008  &  10.8 &   0.98$\pm$  0.12 &  12.4 &   0.56$\pm$  0.09 &   6.8 &          ...      & 1 \\
MS0158.5+0019    & 16-Aug-1996 &  50064004  &   4.3 &   3.97$\pm$  0.33 &  12.4 &   1.65$\pm$  0.12 &   5.6 &          ...      & 1 \\
1ES0229+200      & 16-Jul-2001 &  51472001  &  17.5 &   6.84$\pm$  0.21 &  68.3 &  16.20$\pm$  0.16 &  33.1 &    8.2$\pm$   2.2 & 2 \\
MS0317.0+1834    & 15-Jan-1997 &  50064007  &   5.1 &   3.85$\pm$  0.29 &  14.9 &   3.85$\pm$  0.17 &   4.5 &          ...      & 1 \\
1ES0323+022      & 20-Jan-1998 &  50064011  &   6.2 &   2.81$\pm$  0.23 &  14.4 &   3.83$\pm$  0.17 &   6.8 &          ...      & 1 \\
1ES0347-121      & 10-Jan-1997 &  50064006  &   6.4 &   5.94$\pm$  0.32 &  10.7 &   3.21$\pm$  0.18 &   4.5 &          ...      & 1 \\
1ES0414+009      & 21-Sep-1996 &  50064003  &   0.0 &          ...      &  11.1 &   5.08$\pm$  0.22 &   4.7 &          ...      & 1 \\
1ES0502+675      & 06-Oct-1996 &  50064013  &   0.0 &          ...      &  11.0 &   9.45$\pm$  0.30 &   4.4 &          ...      & 1 \\
1ES0507-040      & 11-Feb-1999 &  50519001  &   9.3 &   4.17$\pm$  0.23 &  20.7 &   6.11$\pm$  0.18 &   9.1 &          ...      & 1 \\
PKS0548-322      & 20-Feb-1999 &  50493003  &   9.5 &  17.57$\pm$  0.44 &  12.5 &  26.11$\pm$  0.46 &   9.4 &          ...      & 4 \\
                 & 26-Feb-1999 &  50493004  &   0.0 &          ...      &   2.1 &  20.77$\pm$  1.03 &   1.0 &          ...      & 1 \\
                 & 07-Apr-1999 &  504930042 &   5.9 &  15.11$\pm$  0.52 &  19.0 &  16.54$\pm$  0.30 &  10.2 &          ...      & 2 \\
MS0737.9+7441    & 29-Oct-1996 &  50064015  &   3.7 &   2.25$\pm$  0.27 &  16.4 &   0.78$\pm$  0.08 &  10.6 &          ...      & 2 \\
B20912+29        & 14-Nov-1997 &  50223004  &   9.4 &   3.89$\pm$  0.22 &  23.9 &   2.96$\pm$  0.12 &  10.7 &          ...      & 1 \\
1ES0927+500      & 25-Nov-1998 &  50519006  &   8.6 &   5.85$\pm$  0.28 &  22.7 &   7.10$\pm$  0.18 &  10.1 &          ...      & 2 \\
1ES1028+511      & 01-May-1997 &  50064009  &   4.6 &  16.32$\pm$  0.61 &  12.6 &   5.46$\pm$  0.21 &   5.0 &   25.0$\pm$   6.9 & 2 \\
RXJ1037.7+5711   & 02-Dec-1998 &  50470003  &  20.9 &   1.71$\pm$  0.11 &  47.0 &   0.97$\pm$  0.06 &  21.6 &          ...      & 1 \\
RXJ1058.6+5628   & 21-May-1998 &  50470002  &  13.9 &   1.32$\pm$  0.12 &  30.8 &   0.36$\pm$  0.05 &  13.8 &          ...      & 1 \\
                 & 23-Nov-1998 &  504700021 &   7.1 &   4.32$\pm$  0.26 &  19.7 &   2.06$\pm$  0.11 &   8.9 &          ...      & 2 \\
1ES1101-232      & 04-Jan-1997 &  50064017  &   6.1 &  24.56$\pm$  0.65 &  13.9 &  20.69$\pm$  0.39 &  10.7 &   21.3$\pm$   5.0 & 3 \\
                 & 19-Jun-1998 &  50726001  &   9.1 &  22.30$\pm$  0.51 &  24.9 &  29.63$\pm$  0.35 &  10.8 &          ...      & 3 \\
MKN421           & 29-Apr-1997 &  50032001  &  11.6 & 168.60$\pm$  1.22 &  23.3 &  51.97$\pm$  0.47 &   9.2 &   24.5$\pm$   4.9 & 4 \\
                 & 30-Apr-1997 &  50032002  &  11.4 & 167.50$\pm$  1.23 &  24.0 &  51.14$\pm$  0.46 &  11.0 &          ...      & 4 \\
                 & 01-May-1997 &  50032003  &  11.2 & 200.30$\pm$  1.35 &  23.7 &  60.70$\pm$  0.51 &  10.9 &          ...      & 4 \\
                 & 02-May-1997 &  50160002  &   4.4 & 243.70$\pm$  2.37 &  11.4 &  74.63$\pm$  0.81 &   5.3 &   28.3$\pm$   6.6 & 4 \\
                 & 03-May-1997 &  50160003  &   4.3 & 155.00$\pm$  1.91 &  11.7 &  44.26$\pm$  0.62 &   5.4 &          ...      & 4 \\
                 & 04-May-1997 &  50160004  &   4.9 & 122.80$\pm$  1.60 &  12.2 &  26.15$\pm$  0.47 &   5.6 &          ...      & 4 \\
                 & 05-May-1997 &  50160005  &   5.0 & 164.40$\pm$  1.84 &  11.9 &  38.75$\pm$  0.57 &   5.5 &          ...      & 4 \\
                 & 07-May-1997 &  50160006  &   6.0 & 139.70$\pm$  1.54 &   0.0 &          ...      &   6.2 &          ...      & 4 \\
                 & 21-Apr-1998 &  50686002  &  23.6 & 335.90$\pm$  1.12 &  29.5 & 220.80$\pm$  0.80 &  13.3 &   45.9$\pm$   4.1 & 4 \\
                 & 23-Apr-1998 &  50686001  &  27.2 & 472.60$\pm$  1.43 &  34.7 & 375.00$\pm$  1.13 &  15.5 &   28.9$\pm$   3.8 & 4 \\
                 & 22-Jun-1998 &  50765001  &  11.3 & 331.70$\pm$  1.73 &  32.5 & 288.70$\pm$  0.94 &   5.7 &          ...      & 4 \\
                 & 04-May-1999 &  50918001  &  63.2 & 262.70$\pm$  0.70 & 111.2 & 141.90$\pm$  0.36 &  57.4 &   11.4$\pm$   1.8 & 6 \\
                 & 26-Apr-2000 &  512820011 &  50.7 & 508.40$\pm$  1.01 &  57.6 & 695.20$\pm$  1.10 &  29.3 &  262.9$\pm$   2.6 & 4 \\
                 & 28-Apr-2000 &  512820012 &  22.5 & 535.50$\pm$  1.55 &  25.7 & 749.20$\pm$  1.71 &  13.6 &  293.5$\pm$   3.9 & 4 \\
                 & 30-Apr-2000 &  512820013 &  60.9 & 522.20$\pm$  0.93 &  33.0 & 677.70$\pm$  1.43 &  32.8 &  205.3$\pm$   2.4 & 6 \\
                 & 09-May-2000 &  512820014 &  63.4 & 444.20$\pm$  0.84 & 131.6 & 561.20$\pm$  0.65 &  65.9 &  203.3$\pm$   1.7 & 6 \\
RXJ1117.1+2014   & 13-Dec-1999 &  50863004  &  13.8 &  16.34$\pm$  0.35 &  42.4 &   7.93$\pm$  0.14 &  23.7 &          ...      & 5 \\
1ES1118+424      & 01-May-1997 &  50064012  &   6.1 &   4.80$\pm$  0.30 &  10.0 &   1.55$\pm$  0.13 &   4.0 &          ...      & 1 \\
1ES1133+704      & 10-Dec-1996 &  50064010  &   5.2 &   8.68$\pm$  0.42 &  18.2 &   2.72$\pm$  0.13 &   7.3 &          ...      & 2 \\
1RXSJ121158+2242 & 27-Dec-1999 &  51000002  &  19.5 &   2.20$\pm$  0.12 &  46.4 &   2.55$\pm$  0.08 &  21.3 &          ...      & 1 \\
                 & 28-Dec-2001 &  51319002  &  20.6 &   4.66$\pm$  0.16 &  56.8 &   8.37$\pm$  0.13 &  24.2 &          ...      & 1 \\
                 & 11-Jan-2002 &  513190021 &   5.1 &   3.84$\pm$  0.30 &  16.4 &   5.94$\pm$  0.20 &   7.1 &          ...      & 1 \\
ON325            & 23-Dec-1998 &  50750001  &  12.5 &   1.01$\pm$  0.11 &  31.4 &   0.85$\pm$  0.07 &  14.8 &          ...      & 1 \\
                 & 10-Jan-2002 &  51472003  &  13.6 &   2.03$\pm$  0.14 &  36.8 &   0.81$\pm$  0.06 &   0.0 &          ...      & 1 \\
1ES1218+304      & 12-Jul-1999 &  50863005  &  10.8 &  21.48$\pm$  0.46 &  42.8 &  18.12$\pm$  0.21 &  20.7 &          ...      & 5 \\
1ES1255+244      & 20-Jun-1998 &  50064001  &   2.5 &  10.47$\pm$  0.67 &   6.9 &  12.13$\pm$  0.43 &   3.3 &          ...      & 1 \\
MS1312.1-4221    & 21-Feb-1997 &  50064016  &   3.5 &   4.23$\pm$  0.37 &   5.6 &   2.45$\pm$  0.22 &   2.4 &          ...      & 1 \\
1RXSJ141756.8+25 & 13-Jul-2000 &  51241001  &  10.6 &  12.60$\pm$  0.35 &  23.7 &  14.89$\pm$  0.26 &  11.1 &   37.6$\pm$   3.9 & 3 \\
                 & 23-Jul-2000 &  512410011 &   3.9 &  10.20$\pm$  0.53 &   7.0 &   9.76$\pm$  0.38 &   3.0 &   32.3$\pm$   7.6 & 3 \\
                 & 27-Jul-2000 &  512410012 &  12.1 &   9.21$\pm$  0.29 &  29.5 &  10.79$\pm$  0.20 &  13.3 &   14.0$\pm$   3.5 & 2 \\
1ES1426+428      & 08-Feb-1999 &  50493006  &  15.5 &  14.76$\pm$  0.32 &  40.8 &  22.36$\pm$  0.24 &  20.4 &   26.4$\pm$   3.1 & 2 \\
MS14588+2249     & 19-Feb-2001 &  51153002  &  22.5 &   2.15$\pm$  0.11 &  47.9 &   0.87$\pm$  0.06 &  21.8 &          ...      & 1 \\
1ES1517+656      & 05-Mar-1997 &  50064014  &   4.5 &  12.50$\pm$  0.54 &  11.1 &   5.19$\pm$  0.22 &   4.0 &          ...      & 3 \\
1ES1533+535      & 13-Feb-1999 &  50519003  &   8.5 &   3.80$\pm$  0.23 &  26.8 &   2.80$\pm$  0.11 &   4.1 &          ...      & 1 \\
1ES1544+820      & 13-Feb-1999 &  50519004  &   8.2 &   2.08$\pm$  0.18 &  23.2 &   1.94$\pm$  0.10 &  10.4 &          ...      & 1 \\
1ES1553+113      & 05-Feb-1998 &  50064005  &   4.5 &  26.38$\pm$  0.78 &  10.6 &  15.30$\pm$  0.39 &   4.7 &          ...      & 3 \\
MKN501           & 07-Apr-1997 &  50377001  &  13.0 & 156.30$\pm$  1.11 &  20.7 & 111.60$\pm$  0.74 &   8.9 &  148.6$\pm$   5.3 & 3 \\
                 & 11-Apr-1997 &  50377002  &  13.1 & 155.80$\pm$  1.10 &  20.4 & 122.20$\pm$  0.77 &   8.7 &  175.8$\pm$   5.4 & 3 \\
                 & 16-Apr-1997 &  50377003  &   9.8 & 238.60$\pm$  1.58 &  17.1 & 258.30$\pm$  1.22 &   7.3 &  556.5$\pm$   6.4 & 4 \\
                 & 28-Apr-1998 &  50529001  &  13.8 & 115.20$\pm$  0.92 &  21.9 & 185.00$\pm$  0.92 &   9.9 &  117.9$\pm$   4.8 & 3 \\
                 & 29-Apr-1998 &  50529002  &  14.9 & 135.90$\pm$  0.96 &  21.4 & 219.40$\pm$  1.01 &   9.7 &  138.0$\pm$   4.9 & 3 \\
                 & 01-May-1998 &  50529003  &  13.4 & 116.20$\pm$  0.94 &  19.0 & 159.80$\pm$  0.92 &   8.4 &   74.0$\pm$   5.1 & 4 \\
                 & 20-Jun-1998 &  50666001  &  17.0 &  43.06$\pm$  0.51 &  25.9 &  55.27$\pm$  0.46 &  11.6 &   31.9$\pm$   3.4 & 3 \\
                 & 29-Jun-1998 &  50666002  &  12.9 &  72.85$\pm$  0.76 &  18.9 & 114.10$\pm$  0.78 &   7.5 &   41.7$\pm$   5.5 & 3 \\
                 & 16-Jul-1998 &  50666003  &  11.4 &  70.81$\pm$  0.80 &  15.9 & 102.70$\pm$  0.80 &   6.9 &   52.1$\pm$   5.7 & 4 \\
                 & 25-Jul-1998 &  50666004  &  25.5 &  78.47$\pm$  0.56 &  31.0 & 104.80$\pm$  0.58 &  14.5 &   32.8$\pm$   3.9 & 4 \\
                 & 10-Jun-1999 &  50944001  & 105.8 &  43.20$\pm$  0.21 & 175.4 &  35.96$\pm$  0.14 &  78.4 &    8.3$\pm$   1.6 & 3 \\
H1722+119        & 24-Aug-2001 &  51472002  &  30.4 &  30.39$\pm$  0.32 &  73.3 &  41.46$\pm$  0.24 &  34.1 &   14.0$\pm$   2.2 & 6 \\
1ES1741+196      & 26-Sep-1998 &  50470001  &  12.7 &   4.78$\pm$  0.21 &  38.7 &   7.11$\pm$  0.14 &  17.4 &          ...      & 1 \\
1ES1959+650      & 04-May-1997 &  50064002  &   2.3 &  15.78$\pm$  0.85 &  12.4 &   7.40$\pm$  0.25 &   5.4 &          ...      & 3 \\
                 & 23-Sep-2001 &  51386001  &   5.8 &  70.50$\pm$  1.11 &   7.2 &  98.84$\pm$  1.18 &   2.9 &          ...      & 4 \\
                 & 28-Sep-2001 &  513860011 &  25.7 &  73.71$\pm$  0.54 &  48.1 & 125.60$\pm$  0.51 &  22.4 &   31.4$\pm$   2.7 & 3 \\

\hline
\end{tabular}
\end{center}
\end{table*}
\begin{table*}
\begin{center}
\scriptsize
\begin{tabular}{lccccccccc}
\hline
\noalign{\smallskip}
\hline
\noalign{\smallskip}
~~~~~Obj. Name     & Obs. Date   & Seq. Num & LECS     &    c/s     & MECS      &      c/s   &  PDS      &     c/s       & Model \\
                   &             &          & (ksec)  & ($10^{-2}$)  &  (ksec)  &  ($10^{-2}$) &  (ksec)  &  ($10^{-2}$) &       \\
~~~~~~~~~~(1)      & (2)         &  (3)     &  (4)     & (5)        &   (6)     &  (7)       & (8)       & (9)           &  (10) \\
\noalign{\smallskip}
\hline
\noalign{\smallskip}

PKS2005-489      & 29-Sep-1996 &  50046001  &   0.0 &          ...      &   9.9 &  35.95$\pm$  0.60 &   7.7 &          ...      & 1 \\
                 & 01-Nov-1998 &  50503002  &  20.7 & 167.50$\pm$  0.91 &  52.5 & 201.40$\pm$  0.62 &  23.4 &   66.4$\pm$   3.0 & 3 \\
PKS2155-304      & 20-Nov-1996 &  50016001  &  36.3 & 132.30$\pm$  0.61 & 106.9 &  33.28$\pm$  0.18 &  43.3 &   13.1$\pm$   2.3 & 4 \\
                 & 22-Nov-1997 &  50160008  &  22.5 & 189.90$\pm$  0.93 &  59.5 & 104.50$\pm$  0.42 &  28.0 &    9.5$\pm$   2.8 & 4 \\
                 & 04-Nov-1999 &  50880001  &  46.1 &  82.97$\pm$  0.43 & 104.0 &  31.43$\pm$  0.18 &  49.1 &          ...      & 4 \\
1ES2344+514      & 03-Dec-1996 &  50320001  &   4.7 &   9.61$\pm$  0.47 &  13.0 &   8.64$\pm$  0.26 &   6.0 &          ...      & 1 \\
                 & 04-Dec-1996 &  50320002  &   5.3 &  10.16$\pm$  0.46 &  13.2 &  10.38$\pm$  0.28 &   6.1 &          ...      & 2 \\
                 & 05-Dec-1996 &  50320003  &   3.5 &  17.09$\pm$  0.72 &   8.0 &  13.89$\pm$  0.42 &   3.8 &          ...      & 1 \\
                 & 07-Dec-1996 &  50320004  &   5.6 &  16.74$\pm$  0.57 &  14.0 &  18.72$\pm$  0.37 &   6.4 &   29.4$\pm$   5.9 & 1 \\
                 & 11-Dec-1996 &  50320005  &   3.0 &   9.82$\pm$  0.60 &  13.0 &  11.71$\pm$  0.30 &   5.8 &          ...      & 1 \\
                 & 26-Jun-1998 &  50760001  &  13.5 &   6.26$\pm$  0.23 &  50.6 &   9.75$\pm$  0.14 &  21.4 &          ...      & 1 \\
                 & 03-Dec-1999 &  50969001  &  28.8 &   5.35$\pm$  0.14 &  90.1 &  10.09$\pm$  0.11 &  45.3 &          ...      & 3 \\
H2356-309        & 21-Jun-1998 &  50493007  &  15.4 &  21.33$\pm$  0.38 &  41.0 &  27.64$\pm$  0.26 &  18.0 &   13.4$\pm$   3.4 & 3 \\

\hline	     
\noalign{\smallskip} 
\multicolumn{9}{c}{LBLs}\\
\noalign{\smallskip}
\hline
\noalign{\smallskip}

PKS0048-09       & 19-Dec-1997 &  500460031 &   4.7 &   1.35$\pm$  0.20 &   9.9 &   1.45$\pm$  0.14 &   4.3 &          ...      & 1 \\
3C66A            & 31-Jan-1999 &  50750003  &   8.9 &   3.11$\pm$  0.20 &  26.4 &   2.36$\pm$  0.11 &  12.7 &          ...      & 4 \\
                 & 29-Jul-2001 &  51372002  &  19.8 &   4.18$\pm$  0.16 &  50.2 &   2.88$\pm$  0.08 &  23.1 &    9.8$\pm$   2.7 & 2 \\
AO0235+164       & 28-Jan-1999 &  50482001  &  25.0 &   0.47$\pm$  0.07 &  29.2 &   0.87$\pm$  0.07 &  25.8 &          ...      & 1 \\
PKS0537-441      & 28-Nov-1998 &  50497002  &  24.1 &   1.71$\pm$  0.10 &  47.4 &   2.44$\pm$  0.08 &  36.7 &          ...      & 1 \\
S50716+71        & 14-Nov-1996 &  50160001  &  16.9 &   2.23$\pm$  0.13 & 122.5 &   0.75$\pm$  0.03 &  47.2 &          ...      & 3 \\
                 & 07-Nov-1998 &  50750002  &   9.5 &   2.39$\pm$  0.18 &  31.3 &   2.86$\pm$  0.13 &  16.2 &   11.5$\pm$   3.5 & 4 \\
                 & 30-Oct-2000 &  51153001  &  19.2 &   3.77$\pm$  0.15 &  43.5 &   3.31$\pm$  0.09 &  22.4 &          ...      & 3 \\
OJ287            & 24-Nov-1997 &  50046002  &   5.2 &   0.72$\pm$  0.15 &  10.7 &   2.58$\pm$  0.17 &   4.5 &          ...      & 1 \\
                 & 20-Nov-2001 &  51336001  &  23.8 &   0.58$\pm$  0.07 &  39.8 &   1.40$\pm$  0.07 &  26.9 &          ...      & 1 \\
PKS1144-379      & 10-Jan-1997 &  50046006  &  10.8 &   0.84$\pm$  0.11 &  23.2 &   0.43$\pm$  0.05 &   9.8 &          ...      & 1 \\
ON231            & 11-May-1998 &  50503001  &  21.1 &   8.01$\pm$  0.20 &  24.9 &   4.14$\pm$  0.14 &  11.6 &   15.8$\pm$   4.3 & 3 \\
                 & 11-Jun-1998 &  505030011 &  17.6 &   5.63$\pm$  0.21 &  31.8 &   3.24$\pm$  0.11 &  16.6 &   14.1$\pm$   3.6 & 3 \\
OQ530            & 12-Feb-1999 &  50482003  &  17.8 &   0.69$\pm$  0.08 &  39.7 &   0.67$\pm$  0.06 &  19.2 &          ...      & 1 \\
                 & 03-Mar-2000 &  50881003  &  19.6 &   0.83$\pm$  0.08 &  26.6 &   1.15$\pm$  0.08 &  10.3 &          ...      & 1 \\
                 & 26-Mar-2000 &  508810031 &  17.2 &   0.72$\pm$  0.09 &  22.7 &   0.95$\pm$  0.08 &  10.4 &          ...      & 1 \\
PKS1519-273      & 01-Feb-1998 &  50046007  &   9.4 &   0.78$\pm$  0.12 &  26.9 &   0.66$\pm$  0.07 &  12.2 &          ...      & 1 \\
S51803+784       & 28-Sep-1998 &  50482004  &  18.7 &   0.77$\pm$  0.09 &  40.5 &   2.10$\pm$  0.08 &  16.4 &          ...      & 1 \\
3C371            & 22-Sep-1998 &  50482002  &  13.3 &   0.96$\pm$  0.10 &  33.9 &   1.90$\pm$  0.09 &  15.1 &          ...      & 1 \\
4C56.27          & 11-Oct-1997 &  50046005  &   4.2 &   0.93$\pm$  0.19 &  13.4 &   1.15$\pm$  0.11 &  59.8 &          ...      & 1 \\
BLLAC            & 08-Nov-1997 &  50046004  &  12.5 &   4.63$\pm$  0.21 &  13.8 &  10.97$\pm$  0.29 &   8.3 &   19.0$\pm$   5.1 & 2 \\
                 & 05-Jun-1999 &  50881001  &  45.3 &   3.94$\pm$  0.10 &  54.4 &   7.13$\pm$  0.12 &  26.9 &   12.1$\pm$   2.7 & 1 \\
                 & 05-Dec-1999 &  50881002  &  17.5 &   3.52$\pm$  0.15 &  54.7 &  12.84$\pm$  0.16 &  24.2 &   11.3$\pm$   2.7 & 1 \\
                 & 26-Jul-2000 &  51165001  &  16.9 &   2.20$\pm$  0.13 &  23.3 &   5.82$\pm$  0.17 &  10.4 &          ...      & 1 \\
                 & 31-Oct-2000 &  511650011 &  24.9 &  18.23$\pm$  0.28 &  33.7 &  23.90$\pm$  0.27 &  18.8 &          ...      & 4 \\

\hline
\noalign{\smallskip}
\multicolumn{9}{c}{FSRQs}\\
\noalign{\smallskip}
\hline
\noalign{\smallskip}

PKS0208-512      & 14-Jan-2001 &  51220002  &  15.9 &   1.55$\pm$  0.12 &  34.3 &   4.92$\pm$  0.13 &  15.3 &   13.7$\pm$   3.3 & 1 \\
NRAO140          & 05-Aug-1999 &  50997001  &  17.5 &   1.83$\pm$  0.12 &  50.0 &   7.36$\pm$  0.13 &  22.9 &   15.2$\pm$   4.7 & 1 \\
PKS0521-365      & 02-Oct-1998 &  50497001  &  18.0 &   4.86$\pm$  0.18 &  41.1 &   9.36$\pm$  0.16 &  19.1 &          ...      & 1 \\
PMN0525-3343     & 27-Feb-2000 &  50849001  &  24.8 &   0.45$\pm$  0.06 &  59.8 &   0.79$\pm$  0.05 &  28.9 &          ...      & 1 \\
PKS0528+134      & 21-Feb-1997 &  50237001  &   6.3 &   0.49$\pm$  0.13 &  14.4 &   1.08$\pm$  0.09 &   6.6 &          ...      & 1 \\
                 & 22-Feb-1997 &  50237002  &   5.0 &          ...      &  13.3 &   1.22$\pm$  0.10 &   5.8 &          ...      & 1 \\
                 & 27-Feb-1997 &  50237003  &   4.4 &          ...      &   7.3 &   0.90$\pm$  0.12 &   3.0 &          ...      & 1 \\
                 & 01-Mar-1997 &  50237004  &   4.3 &          ...      &  13.5 &   1.14$\pm$  0.10 &   6.1 &          ...      & 1 \\
                 & 03-Mar-1997 &  50237005  &   5.1 &          ...      &  14.1 &   0.91$\pm$  0.09 &   6.5 &          ...      & 1 \\
                 & 04-Mar-1997 &  50237006  &   2.8 &          ...      &  11.2 &   1.26$\pm$  0.11 &   4.9 &          ...      & 1 \\
                 & 06-Mar-1997 &  50237007  &   3.1 &          ...      &  12.7 &   1.25$\pm$  0.11 &   5.5 &   19.8$\pm$   6.5 & 1 \\
                 & 11-Mar-1997 &  50237008  &   2.5 &          ...      &  11.5 &   1.05$\pm$  0.10 &   5.3 &   26.6$\pm$   7.0 & 1 \\
WGAJ0546.6-6415  & 01-Oct-1998 &  50769002  &  18.9 &   2.74$\pm$  0.13 &  47.3 &   8.65$\pm$  0.14 &  20.0 &   14.7$\pm$   3.2 & 1 \\
PKS0743-006      & 08-May-2001 &  51143002  &  10.7 &   0.48$\pm$  0.10 &  36.0 &   0.93$\pm$  0.07 &  16.2 &   17.8$\pm$   3.2 & 1 \\
1ES0836+710      & 27-May-1998 &  50497003  &  18.5 &   6.66$\pm$  0.20 &  42.6 &  25.68$\pm$  0.25 &  16.5 &   39.8$\pm$   3.6 & 4 \\
RXJ0909+0354     & 27-Nov-1997 &  50127002  &  10.4 &   0.65$\pm$  0.11 &  29.7 &   1.76$\pm$  0.09 &  13.8 &   14.5$\pm$   4.0 & 1 \\
PKS1127-14       & 04-Jun-1999 &  50850001  &  18.7 &   2.80$\pm$  0.14 &  47.9 &   8.77$\pm$  0.14 &  22.5 &   21.0$\pm$   2.9 & 1 \\
3C273            & 18-Jul-1996 &  50021001  &  11.8 &  31.79$\pm$  0.53 & 129.9 &  35.30$\pm$  0.16 &  60.6 &   83.1$\pm$   2.0 & 3 \\
                 & 13-Jan-1997 &  50237011  &  13.6 &  55.47$\pm$  0.65 &  25.1 &  58.97$\pm$  0.48 &  11.4 &  147.1$\pm$   4.7 & 1 \\
                 & 15-Jan-1997 &  50237012  &  13.3 &  51.86$\pm$  0.63 &  24.0 &  56.80$\pm$  0.49 &  10.8 &  144.9$\pm$   4.9 & 1 \\
                 & 17-Jan-1997 &  50237013  &  12.5 &  49.68$\pm$  0.64 &  27.3 &  55.10$\pm$  0.45 &  12.5 &  130.8$\pm$   4.4 & 1 \\
                 & 22-Jan-1997 &  50237014  &   8.8 &  48.90$\pm$  0.76 &  22.3 &  51.56$\pm$  0.48 &   9.3 &  130.4$\pm$   5.1 & 1 \\
                 & 24-Jun-1998 &  50795004  &  27.9 &  54.87$\pm$  0.45 &  72.2 & 111.00$\pm$  0.39 &  33.8 &  127.3$\pm$   2.6 & 3 \\
                 & 09-Jan-2000 &  50795008  &  34.8 &  58.20$\pm$  0.41 &  85.2 & 119.70$\pm$  0.38 &  42.2 &  135.1$\pm$   2.2 & 3 \\
                 & 13-Jun-2000 &  50795010  &  29.6 &  44.48$\pm$  0.39 &  68.1 &  89.55$\pm$  0.36 &  31.9 &   93.5$\pm$   2.3 & 3 \\
                 & 12-Jun-2001 &  50795011  &  16.9 &  57.99$\pm$  0.59 &  38.4 & 104.80$\pm$  0.52 &  18.5 &  122.7$\pm$   3.1 & 3 \\
3C279            & 13-Jan-1997 &  50237016  &   7.4 &   3.62$\pm$  0.24 &  21.8 &   2.70$\pm$  0.12 &   9.0 &          ...      & 1 \\
                 & 15-Jan-1997 &  50237017  &   8.4 &   3.46$\pm$  0.22 &  23.9 &   2.85$\pm$  0.11 &  10.0 &          ...      & 1 \\
                 & 18-Jan-1997 &  50237018  &   0.5 &          ...      &   2.6 &   2.88$\pm$  0.35 &   1.0 &          ...      & 1 \\
                 & 21-Jan-1997 &  50237019  &   4.2 &   3.47$\pm$  0.31 &  11.7 &   3.23$\pm$  0.17 &   5.1 &          ...      & 1 \\
                 & 23-Jan-1997 &  50237020  &  11.2 &   3.42$\pm$  0.19 &  24.7 &   2.78$\pm$  0.11 &  11.4 &          ...      & 1 \\
GB1428+4217      & 04-Feb-1999 &  50508001  &  27.9 &   1.10$\pm$  0.08 &  90.5 &   2.67$\pm$  0.06 &  46.0 &   14.8$\pm$   2.1 & 1 \\
GB1508+5714      & 29-Mar-1997 &  50127001  &   4.3 &          ...      &   8.8 &   0.75$\pm$  0.11 &   3.9 &          ...      & 1 \\
                 & 01-Feb-1998 &  501270011 &   7.0 &          ...      &  17.5 &   0.51$\pm$  0.08 &   7.8 &          ...      & 1 \\
PKS1510-089      & 03-Aug-1998 &  50497004  &  16.2 &   1.79$\pm$  0.12 &  43.9 &   4.99$\pm$  0.11 &  19.4 &   17.3$\pm$   3.3 & 3 \\
RGBJ1629+4008    & 11-Aug-1999 &  50869001  &  21.2 &   3.28$\pm$  0.14 &  44.9 &   1.40$\pm$  0.07 &  22.8 &          ...      & 1 \\
1ES1641+399      & 19-Feb-1999 &  50727002  &  11.1 &   3.30$\pm$  0.19 &  25.9 &   5.08$\pm$  0.15 &  13.1 &   13.0$\pm$   3.8 & 1 \\
RGBJ1722+2436    & 13-Feb-2000 &  50869002  &  12.4 &   0.71$\pm$  0.10 &  44.1 &   1.08$\pm$  0.06 &  19.7 &          ...      & 1 \\
S41745+62        & 29-Mar-1997 &  50127006  &  58.4 &   0.41$\pm$  0.04 & 143.0 &   0.71$\pm$  0.03 &  59.8 &          ...      & 1 \\
\hline
\end{tabular}
\end{center}
\end{table*}
\begin{table*}
\begin{center}
\scriptsize
\begin{tabular}{lccccccccc}
\hline
\noalign{\smallskip}
\hline
\noalign{\smallskip}
~~~~~Obj. Name     & Obs. Date   & Seq. Num & LECS     &    c/s     & MECS      &      c/s   &  PDS      &     c/s       & Model \\
                   &             &          & (ksec)  & ($10^{-2}$)  &  (ksec)  &  ($10^{-2}$) &  (ksec)  &  ($10^{-2}$) &       \\
~~~~~~~~~~(1)      & (2)         &  (3)     &  (4)     & (5)        &   (6)     &  (7)       & (8)       & (9)           &  (10) \\
\noalign{\smallskip}
\hline
\noalign{\smallskip}

S52116+81        & 29-Apr-1998 &  50769001  &  13.4 &   7.54$\pm$  0.25 &  28.9 &  16.71$\pm$  0.24 &  13.3 &   15.9$\pm$   4.0 & 1 \\
                 & 12-Oct-1998 &  507690011 &   5.7 &   5.68$\pm$  0.33 &  19.7 &  11.88$\pm$  0.25 &  10.5 &   13.7$\pm$   4.5 & 1 \\
PKS2126-158      & 24-May-1999 &  51003001  &  79.9 &   3.62$\pm$  0.07 & 108.0 &  11.07$\pm$  0.10 &  51.6 &   10.5$\pm$   2.0 & 4 \\
PKS2134+004      & 25-Nov-2000 &  51143001  &  25.9 &   1.52$\pm$  0.09 &  75.1 &   2.14$\pm$  0.06 &  33.1 &          ...      & 1 \\
PKS2149-306      & 31-Oct-1997 &  50264001  &  17.9 &   2.74$\pm$  0.14 &  39.4 &   7.46$\pm$  0.15 &  16.7 &   13.0$\pm$   3.7 & 1 \\
PKS2223+210      & 12-Nov-1997 &  50127003  &  14.2 &   0.81$\pm$  0.10 &  18.4 &   2.09$\pm$  0.12 &  10.8 &          ...      & 1 \\
PKS2223-05       & 10-Nov-1997 &  50181006  &   9.4 &   0.79$\pm$  0.12 &  16.2 &   1.32$\pm$  0.11 &   6.9 &          ...      & 1 \\
H2230+114        & 11-Nov-1997 &  50237021  &  11.0 &   2.03$\pm$  0.15 &  23.8 &   5.45$\pm$  0.16 &  11.6 &          ...      & 1 \\
                 & 13-Nov-1997 &  50237022  &  10.3 &   2.91$\pm$  0.19 &  22.3 &   5.92$\pm$  0.17 &  10.2 &          ...      & 1 \\
                 & 16-Nov-1997 &  50237023  &  13.2 &   1.91$\pm$  0.14 &  25.7 &   5.84$\pm$  0.16 &  12.1 &   13.7$\pm$   4.3 & 1 \\
                 & 18-Nov-1997 &  50237024  &   7.0 &   2.12$\pm$  0.20 &  12.2 &   5.63$\pm$  0.22 &   5.4 &          ...      & 1 \\
                 & 21-Nov-1997 &  50237025  &   9.5 &   2.49$\pm$  0.18 &  19.4 &   5.98$\pm$  0.18 &   8.6 &          ...      & 1 \\
PKS2243-123      & 18-Nov-1998 &  50727005  &  10.2 &   1.32$\pm$  0.14 &  27.5 &   1.89$\pm$  0.09 &  13.2 &          ...      & 1 \\
H2251+158        & 05-Jun-2000 &  51220001  &  17.7 &   2.94$\pm$  0.14 &  48.5 &  10.92$\pm$  0.15 &  22.4 &   20.3$\pm$   2.8 & 1 \\

\hline
\end{tabular}
\end{center}
{\bf Columns}: {\bf 1}=Object Name; {\bf 2}=Observation Date; {\bf 3}=Sequence 
Number of the observation; {\bf 4}=Exposure time for LECS; 
{\bf 5}=Counts rate for LECS in the energy range 0.1--2 keV; {\bf 6}=Exposure time for 
MECS (for the observations made before May 1997, MECS2 count-rates are considered); 
{\bf 7}=Counts rate for MECS in the energy range 2--10 keV; {\bf 8}=Exposure time for PDS; 
{\bf 9}=Counts rate for PDS in the energy range 13--50 keV. Less than  
3$\sigma$ detections are omitted; {\bf 10}=Model used for the best fit: 
1=Power law with absorption fixed at Galactic value; 2=Power law with free absorption; 
3=Broken power law with absorption fixed at Galactic value; 4=Continuously curved parabola
with absorption fixed at Galactic value; 5=Broken power law with free absorption;
6=Continuously curved parabola with free absorption.
\end{table*}


\clearpage
\begin{table*}
\caption{Sources with power law as best-fit model }
\label{tab:bestfit}
\begin{center}
\scriptsize
\begin{tabular}{lccccccc}
\hline
\noalign{\smallskip}
\hline
\noalign{\smallskip}
&&\multicolumn{2}{c}{$N_{\rm H}$ FIXED }&\multicolumn{3}{c}{$N_{\rm H}$ FREE}&\\
\cline{3-4} \cline{5-7}\\
~~~~~Obj. Name     & Obs. Date  & $\Gamma $ & $\chi^{2}_{r}$/d.o.f.& $N_{\rm H}$ &$\Gamma $  & $\chi^{2}_{r}$/d.o.f.  & Flux   \\
~~~~~~~~~~(1)      & (2)        &  (3)   &  (4)                & (5)             &   (6)                  &  (7)      & (8)     \\
\noalign{\smallskip}
\hline
\noalign{\smallskip}
\multicolumn{8}{c}{HBLs}\\
\noalign{\smallskip}
\hline
\noalign{\smallskip}
1ES0033+595        & 18-Dec-1999  &  &   & $ 35.5^{+  8.3}_{-  7.6}$ & $2.06^{+0.03}_{-0.03}$ & 1.22/155 &   58.9  \\ 
1ES0120+340        & 03-Jan-1999  &  &   & $  9.3^{+  3.3}_{-  2.4}$ & $2.13^{+0.06}_{-0.05}$ & 1.07/99  &   17.1  \\	
                   & 02-Feb-1999  &  &   & $  5.6^{+  6.5}_{-  2.5}$ & $2.33^{+0.09}_{-0.08}$ & 0.84/39  &   13.3  \\	
RXJ0136.5+3905     & 09-Jan-2001  &  &   & $  7.9^{+  3.3}_{-  2.5}$ & $2.58^{+0.07}_{-0.06}$ & 1.27/99  &   10.8  \\	
1ES0145+138        & 30-Dec-1997  & $1.99^{+0.60}_{-0.47}$ & 0.91/24  &  &  &   &    0.6  \\	
MS0158.5+0019      & 16-Aug-1996  & $2.28^{+0.27}_{-0.17}$ & 0.86/25  &  &  &   &    2.9  \\	
1ES0229+200        & 16-Jul-2001  &  &   & $ 10.0^{+  5.1}_{-  4.1}$ & $1.99^{+0.05}_{-0.04}$ & 1.03/99  &   14.9  \\	
MS0317.0+1834      & 15-Jan-1997  & $2.00^{+0.13}_{-0.13}$ & 0.98/59  &  &  &   &    7.1  \\	
1ES0323+022        & 20-Jan-1998  & $2.19^{+0.27}_{-0.17}$ & 0.60/32  &  &  &   &    3.5  \\	
1ES0347-121        & 10-Jan-1997  & $2.03^{+0.11}_{-0.14}$ & 1.04/54  &  &  &   &    6.2  \\	
1ES0414+009        & 21-Sep-1996  & $2.53^{+0.09}_{-0.08}$ & 0.94/54  &  &  &   &    8.7  \\	
1ES0502+675        & 06-Oct-1996  & $2.30^{+0.23}_{-0.22}$ & 1.13/97  &  &  &   &   19.2  \\	
1ES0507-040        & 11-Feb-1999  & $2.14^{+0.14}_{-0.11}$ & 0.78/36  &  &  &   &    5.9  \\	
PKS0548-322        & 26-Feb-1999  & $2.28^{+0.43}_{-0.22}$ & 0.61/16  &  &  &   &   17.8  \\	
                   & 07-Apr-1999  &  &   & $  3.1^{+  0.8}_{-  0.8}$ & $2.25^{+0.08}_{-0.07}$ & 0.92/35  &   15.7  \\	
MS0737.9+7441      & 29-Oct-1996  &  &   & $ 27.3^{+ 53.5}_{- 18.0}$ & $2.70^{+0.41}_{-0.55}$ & 0.95/16  &    1.4  \\	
B20912+29          & 14-Nov-1997  & $2.05^{+0.11}_{-0.11}$ & 1.01/23  &  &  &   &    2.8  \\	
1ES0927+500        & 25-Nov-1998  &  &   & $  3.3^{+  1.5}_{-  1.1}$ & $2.11^{+0.10}_{-0.11}$ & 0.96/33  &    6.8  \\	
1ES1028+511        & 01-May-1997  &  &   & $  4.7^{+  1.5}_{-  1.3}$ & $2.42^{+0.11}_{-0.11}$ & 0.91/91  &   10.1  \\	
RXJ1037.7+5711     & 02-Dec-1998  & $2.23^{+0.19}_{-0.20}$ & 0.94/24  &  &  &   &    0.9  \\	
RXJ1058.6+5628     & 21-May-1998  & $2.26^{+0.60}_{-0.54}$ & 1.08/21  &  &  &   &    0.3  \\	
                   & 23-Nov-1998  &  &   & $  2.3^{+  1.3}_{-  0.8}$ & $2.72^{+0.25}_{-0.24}$ & 0.79/33  &    1.7  \\	
1ES1118+424        & 01-May-1997  & $2.19^{+0.15}_{-0.14}$ & 0.90/38  &  &  &   &    3.1  \\	
1ES1133+704        & 10-Dec-1996  &  &   & $  2.0^{+  0.9}_{-  0.8}$ & $2.55^{+0.18}_{-0.16}$ & 0.85/39  &    5.0  \\	
RXJ1211.9+2242     & 27-Dec-1999  & $1.93^{+0.10}_{-0.10}$ & 1.35/24  &  &  &   &    2.5  \\	
                   & 28-Dec-2001  & $1.93^{+0.06}_{-0.06}$ & 1.04/90  &  &  &   &    7.6  \\	
                   & 11-Jan-2002  & $2.05^{+0.12}_{-0.12}$ & 1.00/40  &  &  &   &    5.6  \\	
ON325              & 23-Dec-1998  & $2.55^{+0.31}_{-0.27}$ & 1.32/25  &  &  &   &    0.6  \\	
                   & 10-Jan-2002  & $2.49^{+0.18}_{-0.17}$ & 0.90/33  &  &  &   &    0.6  \\	
1ES1255+244        & 20-Jun-1998  & $1.95^{+0.11}_{-0.11}$ & 0.79/24  &  &  &   &   11.7  \\	
MS1312.1-4221      & 21-Feb-1997  & $2.31^{+0.25}_{-0.25}$ & 0.80/16  &  &  &   &    4.6  \\	
1RXSJ141756.8+25   & 27-Jul-2000  &  &   & $  2.3^{+  0.8}_{-  0.7}$ & $2.20^{+0.07}_{-0.08}$ & 1.27/96  &    9.3  \\	
1ES1426+428        & 08-Feb-1999  &  &   & $  0.8^{+  0.4}_{-  0.4}$ & $1.94^{+0.04}_{-0.05}$ & 1.12/142 &   20.3  \\	
MS14588+2249       & 19-Feb-2001  & $2.62^{+0.23}_{-0.12}$ & 1.24/24  &  &  &   &    0.7  \\	
1ES1533+535        & 13-Feb-1999  & $2.14^{+0.12}_{-0.06}$ & 1.22/69  &  &  &   &    2.6  \\
1ES1544+820        & 13-Feb-1999  & $2.64^{+0.21}_{-0.10}$ & 0.93/39  &  &  &   &    1.5  \\
1ES1741+196        & 26-Sep-1998  & $2.02^{+0.08}_{-0.07}$ & 1.14/35  &  &  &   &    6.9  \\
PKS2005-489        & 29-Sep-1996  & $2.36^{+0.05}_{-0.05}$ & 1.06/118 &  &  &   &   58.7  \\
1ES2344+514        & 03-Dec-1996  & $2.05^{+0.09}_{-0.08}$ & 0.93/39  &  &  &   &   17.2  \\
                   & 04-Dec-1996  &  &   & $ 13.3^{+  6.8}_{-  6.0}$ & $2.05^{+0.08}_{-0.09}$ & 1.10/134 &   18.8  \\
                   & 05-Dec-1996  & $2.12^{+0.09}_{-0.08}$ & 1.59/51  &  &  &   &   27.1  \\
                   & 07-Dec-1996  & $1.82^{+0.05}_{-0.05}$ & 0.96/230 &  &  &   &   35.3  \\
                   & 11-Dec-1996  & $2.17^{+0.03}_{-0.03}$ & 1.15/144 &  &  &   &   20.4  \\
                   & 26-Jun-1998  & $2.32^{+0.06}_{-0.06}$ & 0.98/93  &  &  &   &    8.5  \\
                   						      
\noalign{\smallskip}						      
\hline     
\noalign{\smallskip}
\multicolumn{8}{c}{LBLs}\\
\noalign{\smallskip}
\hline
\noalign{\smallskip}
PKS0048-09         & 19-Dec-1997  & $1.96^{+0.01}_{-0.01}$ & 1.11/13  &  &  &   &    1.4  \\
3C66A              & 29-Jul-2001  &  &   & $ 14.4^{+  7.6}_{-  7.4}$ & $2.40^{+0.15}_{-0.13}$ & 1.49/34  &    2.7  \\ 
AO0235+164         & 28-Jan-1999  & $2.13^{+0.33}_{-0.32}$ & 1.15/26  &  &  &   &    0.7  \\
PKS0537-441        & 28-Nov-1998  & $1.80^{+0.12}_{-0.12}$ & 0.66/24  &  &  &   &    2.4  \\
OJ287              & 24-Nov-1997  & $1.95^{+0.30}_{-0.29}$ & 0.88/24  &  &  &   &    2.4  \\
                   & 20-Nov-2001  & $1.64^{+0.22}_{-0.20}$ & 0.93/51  &  &  &   &    1.3  \\
PKS1144-379        & 10-Jan-1997  & $1.83^{+0.24}_{-0.44}$ & 0.88/26  &  &  &   &    1.1  \\
OQ530              & 12-Feb-1999  & $1.92^{+0.29}_{-0.29}$ & 0.86/26  &  &  &   &    0.6  \\
                   & 03-Mar-2000  & $1.53^{+0.29}_{-0.32}$ & 1.06/39  &  &  &   &    1.1  \\
                   & 26-Mar-2000  & $1.88^{+0.30}_{-0.15}$ & 0.83/32  &  &  &   &    0.8  \\
PKS1519-273        & 01-Feb-1998  & $2.25^{+0.44}_{-0.62}$ & 1.46/16  &  &  &   &    0.5  \\
S51803+784         & 28-Sep-1998  & $1.53^{+0.16}_{-0.16}$ & 1.04/22  &  &  &   &    2.2  \\
3C371              & 22-Sep-1998  & $1.74^{+0.10}_{-0.20}$ & 1.14/51  &  &  &   &    1.7  \\
4C56.27            & 11-Oct-1997  & $1.84^{+0.53}_{-0.48}$ & 0.97/20  &  &  &   &    1.1  \\
\hline
\end{tabular}
\end{center}
\end{table*}
\begin{table*}
\begin{center}
\scriptsize
\begin{tabular}{lccccccc}
\hline
\noalign{\smallskip}
\hline
\noalign{\smallskip}
&&\multicolumn{2}{c}{$N_{\rm H}$ FIXED }&\multicolumn{3}{c}{$N_{\rm H}$ FREE}&\\ 
\cline{3-4} \cline{5-7}\\
~~~~~Obj. Name     & Obs. Date  & $\Gamma $ & $\chi^{2}_{r}$/d.o.f.& $N_{\rm H}$ &$\Gamma $  & $\chi^{2}_{r}$/d.o.f.  & Flux   \\
~~~~~~~~~~(1)      & (2)        &  (3)   &  (4)                & (5)             &   (6)                  &  (7)      & (8)     \\
\noalign{\smallskip}
\hline
\noalign{\smallskip}
BLLAC              & 08-Nov-1997  &  &   & $ 11.0^{+  8.4}_{-  6.9}$ & $1.88^{+0.06}_{-0.12}$ & 1.11/33  &   11.1  \\
                   & 05-Jun-1999  & $2.02^{+0.07}_{-0.07}$ & 1.12/93  &  &  &   &    6.5  \\
                   & 05-Dec-1999  & $1.59^{+0.05}_{-0.05}$ & 0.99/93  &  &  &   &   12.7  \\
                   & 26-Jul-2000  & $1.88^{+0.12}_{-0.12}$ & 1.21/24  &  &  &   &    5.8  \\
\noalign{\smallskip}
\hline
\noalign{\smallskip}
\multicolumn{8}{c}{FSRQs}\\
\noalign{\smallskip}
\hline
\noalign{\smallskip}
PKS0208-512        & 14-Jan-2001  & $1.64^{+0.10}_{-0.10}$ & 1.33/93  &  &  &   &    4.5  \\
NRAO140            & 05-Aug-1999  & $1.57^{+0.07}_{-0.07}$ & 1.26/93  &  &  &   &    7.2  \\
PKS0521-365        & 02-Oct-1998  & $1.72^{+0.06}_{-0.06}$ & 1.00/147 &  &  &   &    8.7  \\
PMN0525-3343       & 27-Feb-2000  & $1.58^{+0.23}_{-0.24}$ & 0.99/52  &  &  &   &    0.7  \\
PKS0528+134        & 21-Feb-1997  & $1.28^{+0.26}_{-0.26}$ & 0.94/27  &  &  &   &    2.7  \\
                   & 22-Feb-1997  & $1.26^{+0.28}_{-0.30}$ & 0.45/24  &  &  &   &    2.9  \\
                   & 27-Feb-1997  & $1.40^{+0.55}_{-0.57}$ & 0.94/14  &  &  &   &    2.1  \\
                   & 01-Mar-1997  & $1.34^{+0.29}_{-0.29}$ & 1.14/25  &  &  &   &    3.3  \\
                   & 03-Mar-1997  & $1.15^{+0.30}_{-0.31}$ & 1.25/24  &  &  &   &    1.9  \\
                   & 04-Mar-1997  & $1.28^{+0.33}_{-0.34}$ & 0.76/20  &  &  &   &    2.6  \\
                   & 06-Mar-1997  & $1.12^{+0.26}_{-0.26}$ & 1.09/23  &  &  &   &    2.9  \\
                   & 11-Mar-1997  & $1.62^{+0.36}_{-0.32}$ & 1.48/21  &  &  &   &    2.2  \\
WGAJ0546.6-6415    & 01-Oct-1998  & $1.59^{+0.06}_{-0.06}$ & 1.87/93  &  &  &   &    8.2  \\
PKS0743-006        & 08-May-2001  & $1.91^{+0.34}_{-0.30}$ & 0.81/35  &  &  &   &    0.9  \\
RGBJ0909+039       & 27-Nov-1997  & $1.16^{+0.20}_{-0.20}$ & 0.73/42  &  &  &   &    1.9  \\
PKS1127-14         & 04-Jun-1999  & $1.42^{+0.06}_{-0.06}$ & 0.94/93  &  &  &   &    8.7  \\
3C273              & 13-Jan-1997  & $1.56^{+0.02}_{-0.02}$ & 0.91/231 &  &  &   &  117.0  \\
                   & 15-Jan-1997  & $1.58^{+0.02}_{-0.02}$ & 1.19/231 &  &  &   &  112.0  \\
                   & 17-Jan-1997  & $1.62^{+0.02}_{-0.02}$ & 1.35/231 &  &  &   &  108.0  \\
                   & 22-Jan-1997  & $1.55^{+0.02}_{-0.02}$ & 0.97/231 &  &  &   &  103.0  \\
3C279              & 13-Jan-1997  & $1.67^{+0.12}_{-0.12}$ & 0.98/39  &  &  &   &    5.6  \\
                   & 15-Jan-1997  & $1.70^{+0.11}_{-0.11}$ & 0.75/50  &  &  &   &    5.8  \\
                   & 18-Jan-1997  & $1.43^{+0.47}_{-0.47}$ & 1.00/13  &  &  &   &    7.4  \\
                   & 21-Jan-1997  & $1.53^{+0.16}_{-0.16}$ & 0.79/48  &  &  &   &    6.4  \\
                   & 23-Jan-1997  & $1.59^{+0.10}_{-0.10}$ & 1.18/50  &  &  &   &    5.7  \\
GB1428+4217        & 04-Feb-1999  & $1.53^{+0.09}_{-0.10}$ & 1.15/35  &  &  &   &    2.8  \\
GB1508+5714        & 29-Mar-1997  & $0.65^{+0.83}_{-0.85}$ & 0.84/14  &  &  &   &    1.3  \\
                   & 01-Feb-1998  & $1.27^{+0.84}_{-0.79}$ & 1.33/18  &  &  &   &    0.5  \\
RGBJ1629+4008      & 11-Aug-1999  & $2.55^{+0.09}_{-0.09}$ & 1.19/35  &  &  &   &    1.3  \\
1ES1641+399        & 19-Feb-1999  & $1.71^{+0.10}_{-0.05}$ & 1.92/35  &  &  &   &    5.1  \\
RGBJ1722+2436      & 13-Feb-2000  & $1.54^{+0.24}_{-0.12}$ & 1.00/44  &  &  &   &    1.0  \\
S41745+62          & 29-Mar-1997  & $1.64^{+0.17}_{-0.18}$ & 0.94/97  &  &  &   &    0.8  \\
S52116+81          & 29-Apr-1998  & $1.72^{+0.05}_{-0.05}$ & 1.23/93  &  &  &   &   15.8  \\
                   & 12-Oct-1998  & $1.83^{+0.09}_{-0.09}$ & 1.25/35  &  &  &   &   11.9  \\
PKS2134+004        & 25-Nov-2000  & $1.71^{+0.06}_{-0.12}$ & 1.69/22  &  &  &   &    2.2  \\
PKS2149-306        & 31-Oct-1997  & $1.38^{+0.08}_{-0.08}$ & 1.13/32  &  &  &   &    8.1  \\
PKS2223+210        & 12-Nov-1997  & $1.31^{+0.12}_{-0.26}$ & 0.87/37  &  &  &   &    2.2  \\
PKS2223-05         & 10-Nov-1997  & $1.87^{+0.37}_{-0.35}$ & 1.17/15  &  &  &   &    1.3  \\
H2230+114          & 11-Nov-1997  & $1.57^{+0.12}_{-0.12}$ & 0.87/35  &  &  &   &    5.6  \\
                   & 13-Nov-1997  & $1.50^{+0.11}_{-0.11}$ & 0.95/35  &  &  &   &    6.2  \\
                   & 16-Nov-1997  & $1.47^{+0.11}_{-0.10}$ & 0.97/35  &  &  &   &    6.2  \\
                   & 18-Nov-1997  & $1.49^{+0.17}_{-0.16}$ & 1.27/41  &  &  &   &    5.5  \\
                   & 21-Nov-1997  & $1.50^{+0.13}_{-0.13}$ & 1.16/35  &  &  &   &    6.2  \\
PKS2243-123        & 18-Nov-1998  & $1.67^{+0.12}_{-0.22}$ & 0.97/44  &  &  &   &    1.8  \\
H2251+158          & 05-Jun-2000  & $1.34^{+0.06}_{-0.05}$ & 1.31/93  &  &  &   &   11.2  \\
\hline
\end{tabular}
\end{center}
{\bf Columns}: {\bf 1}=Object Name; {\bf 2}=Observation Date; {\bf 3}=Photon spectral 
index; {\bf 4}=Reduced $\chi^{2}$ with degrees of freedom. Columns 3 and 4 refer
to fits obtained with intrinsic absorption fixed to the Galactic value; {\bf 5}=Intrinsic 
absorption (in units of $10^{20}{\rm cm}^{-2}$); {\bf 6}=Photon spectral index;
{\bf 7}=Reduced $\chi^{2}$ with degrees of freedom. Columns 5 to 7 refer
to fits obtained with intrinsic absorption free to vary; 
{\bf 8}=X--ray unabsorbed flux in 
the energy range 2--10 keV (in units of $10^{-12}$ erg cm$^{-2}$ s$^{-1}$).
\end{table*}

\clearpage
\begin{table*}
\caption{Spectra best-fitted by a curved model }
\label{tab:bestfit2}
\begin{center}
\rotate{
\scriptsize
\begin{tabular}{lccccccccccc}
\hline
\noalign{\smallskip}
\hline
\noalign{\smallskip}
&&\multicolumn{4}{c}{$N_{\rm H}$ FIXED }&\multicolumn{5}{c}{$N_{\rm H}$ FREE}&\\ 
\cline{4-7} \cline{8-12}\\
~~~~~Obj. Name     & Obs. Date   & $\chi^{2}_{r}$/d.o.f. & $\Gamma_{1}$   & $E_{break}$   & $\Gamma_{2}$  & $\chi^{2}_{r}$/d.o.f.& $N_{\rm H}$ &$\Gamma_{1}$   & $E_{break}$   & $\Gamma_{2}$  & Flux  \\
~~~~~~~~~~(1)      & (2)       &  (3)   &  (4)      & (5)      &   (6)       &  (7)      & (8)     & (9)    & (10)    & (11)   & (12)    \\
\noalign{\smallskip}
\hline
\noalign{\smallskip}
\multicolumn{12}{c}{HBLs}\\
\noalign{\smallskip}
\hline
\noalign{\smallskip}
PKS0548-322        & 20-Feb-1999  & 1.21/91  & $1.66^{+0.13}_{-0.09}$ & $ 0.58^{+99.42}_{- 0.58}$ & $2.45^{+0.40}_{-0.15}$ &   &  &  &  &  &  22.3  \\	
1ES1101-232        & 04-Jan-1997  & 0.91/254 & $1.73^{+0.13}_{-0.09}$ & $ 3.45^{+ 0.66}_{- 0.52}$ & $2.19^{+0.14}_{-0.11}$ &   &  &  &  &  &  36.8  \\	
                   & 19-Jun-1998  & 1.02/88  & $1.77^{+0.16}_{-0.24}$ & $ 1.25^{+ 0.43}_{- 0.25}$ & $2.28^{+0.05}_{-0.05}$ &   &  &  &  &  &  25.6  \\	
Mkn421             & 29-Apr-1997  & 1.14/229 & $2.26^{+0.04}_{-0.04}$ & $ 0.14^{+ 0.59}_{- 0.12}$ & $3.05^{+0.09}_{-0.07}$ &   &  &  &  &  &  84.5  \\	
                   & 30-Apr-1997  & 0.79/229 & $2.25^{+0.04}_{-0.04}$ & $ 0.25^{+ 0.88}_{- 0.22}$ & $3.10^{+0.09}_{-0.08}$ &   &  &  &  &  &  83.2  \\	
                   & 01-May-1997  & 1.10/229 & $2.23^{+0.04}_{-0.03}$ & $ 0.16^{+ 0.57}_{- 0.15}$ & $3.02^{+0.08}_{-0.07}$ &   &  &  &  &  &  99.2  \\	
                   & 02-May-1997  & 1.09/229 & $2.22^{+0.05}_{-0.05}$ & $ 0.22^{+ 1.31}_{- 0.21}$ & $3.08^{+0.11}_{-0.09}$ &   &  &  &  &  & 121.0  \\	
                   & 03-May-1997  & 1.07/229 & $2.21^{+0.06}_{-0.04}$ & $ 3.29^{+ 5.95}_{- 1.85}$ & $3.34^{+0.32}_{-0.22}$ &   &  &  &  &  &  70.3  \\	
                   & 04-May-1997  & 1.07/229 & $2.49^{+0.06}_{-0.08}$ & $ 0.16^{+ 2.59}_{- 0.15}$ & $3.38^{+0.20}_{-0.13}$ &   &  &  &  &  &  40.1  \\	
                   & 05-May-1997  & 1.27/216 & $2.39^{+0.02}_{-0.02}$ & $ 0.18^{+ 0.15}_{- 0.09}$ & $3.22^{+0.05}_{-0.05}$ &   &  &  &  &  &  60.7  \\	
                   & 07-May-1997  & 0.85/42  & $2.32^{+0.07}_{-0.06}$ & $ 2.02^{+ 0.93}_{- 1.33}$ & $3.63^{+0.37}_{-0.83}$ &   &  &  &  &  &  37.8  \\	
                   & 21-Apr-1998  & 0.95/146 & $2.17^{+0.02}_{-0.02}$ & $ 1.14^{+ 1.40}_{- 0.63}$ & $3.00^{+0.05}_{-0.05}$ &   &  &  &  &  & 178.0  \\	
                   & 23-Apr-1998  & 1.12/146 & $2.02^{+0.02}_{-0.02}$ & $ 1.18^{+ 0.15}_{- 0.13}$ & $2.64^{+0.02}_{-0.02}$ &   &  &  &  &  & 312.0  \\	
                   & 22-Jun-1998  & 1.45/133 & $2.05^{+0.02}_{-0.02}$ & $ 0.43^{+ 0.92}_{- 0.32}$ & $2.74^{+0.05}_{-0.05}$ &   &  &  &  &  & 240.0  \\	
                   & 04-May-1999  & & & & & 1.89/145 & $  0.6^{+ 0.1}_{- 0.1}$ & $2.36^{+0.01}_{-0.02}$ & $ 1.80^{+ 0.20}_{- 0.16}$ & $3.02^{+0.03}_{-0.02}$ &  111.0  \\	
                   & 26-Apr-2000  & 2.58/146 & $1.73^{+0.01}_{-0.01}$ & $ 1.39^{+ 0.30}_{- 0.30}$ & $2.22^{+0.01}_{-0.01}$ &   &  &  &  &  & 619.0  \\	
                   & 28-Apr-2000  & 1.93/145 & $1.73^{+0.01}_{-0.01}$ & $ 1.45^{+ 1.38}_{- 0.75}$ & $2.23^{+0.02}_{-0.02}$ &   &  &  &  &  & 669.0  \\	
                   & 30-Apr-2000  & & & & & 2.82/145 & $  0.3^{+ 1.4}_{- 0.3}$ & $1.79^{+0.01}_{-0.01}$ & $ 4.57^{+ 0.78}_{- 0.77}$ & $2.35^{+0.02}_{-0.02}$ &  598.0  \\	
                   & 09-May-2000  & & & & & 3.95/140 & $  0.3^{+ 0.9}_{- 0.3}$ & $1.84^{+0.01}_{-0.01}$ & $ 2.12^{+ 0.14}_{- 0.14}$ & $2.19^{+0.01}_{-0.01}$ &  497.0  \\	
RXJ1117.1+2014     & 13-Dec-1999  & & & & & 1.12/83  & $  1.5^{+ 0.7}_{- 0.8}$ & $2.49^{+0.16}_{-0.22}$ & $ 1.52^{+ 0.90}_{- 0.39}$ & $2.98^{+0.09}_{-0.08}$ &    6.1  \\	
1ES1218+304        & 12-Jul-1999  & & & & & 0.77/93  & $  1.5^{+ 0.7}_{- 0.8}$ & $2.21^{+0.13}_{-0.13}$ & $ 2.57^{+ 1.59}_{- 1.18}$ & $2.57^{+0.26}_{-0.06}$ &   14.9  \\	
1RXSJ141756.8+25   & 13-Jul-2000  & 0.95/96  & $1.79^{+0.07}_{-0.08}$ & $ 3.00^{+ 0.48}_{- 0.68}$ & $2.40^{+0.14}_{-0.14}$ &   &  &  &  &  &  12.6  \\	
                   & 23-Jul-2000  & 0.86/35  & $1.74^{+0.22}_{-0.18}$ & $ 3.10^{+ 0.54}_{- 0.66}$ & $2.60^{+0.34}_{-0.27}$ &   &  &  &  &  &   8.7  \\	
1ES1517+656        & 05-Mar-1997  & 1.45/39  & $1.40^{+0.19}_{-0.24}$ & $ 1.23^{+ 0.19}_{- 0.18}$ & $2.60^{+0.15}_{-0.07}$ &   &  &  &  &  &   9.5  \\
1ES1553+113        & 05-Feb-1998  & 0.98/30  & $1.96^{+0.19}_{-0.21}$ & $ 1.01^{+ 0.46}_{- 0.18}$ & $2.79^{+0.12}_{-0.11}$ &   &  &  &  &  &  13.7  \\
Mkn501             & 07-Apr-1997  & 0.97/234 & $1.50^{+0.04}_{-0.03}$ & $ 1.08^{+ 0.09}_{- 0.08}$ & $1.92^{+0.02}_{-0.02}$ &   &  &  &  &  & 208.0  \\
                   & 11-Apr-1997  & 1.04/233 & $1.48^{+0.02}_{-0.02}$ & $ 1.08^{+ 0.09}_{- 0.08}$ & $1.92^{+0.02}_{-0.02}$ &   &  &  &  &  & 232.0  \\
                   & 16-Apr-1997  & 1.00/229 & $1.41^{+0.02}_{-0.02}$ & $ 2.88^{+ 1.33}_{- 1.02}$ & $1.69^{+0.02}_{-0.02}$ &   &  &  &  &  & 510.0  \\
                   & 28-Apr-1998  & 1.19/142 & $1.54^{+0.04}_{-0.04}$ & $ 1.14^{+ 0.13}_{- 0.14}$ & $1.85^{+0.02}_{-0.02}$ &   &  &  &  &  & 172.0  \\
                   & 29-Apr-1998  & 1.02/140 & $1.51^{+0.03}_{-0.03}$ & $ 1.26^{+ 0.12}_{- 0.11}$ & $1.85^{+0.02}_{-0.02}$ &   &  &  &  &  & 204.0  \\
                   & 01-May-1998  & 1.09/143 & $1.70^{+0.04}_{-0.03}$ & $ 2.65^{+ 4.27}_{- 1.62}$ & $2.24^{+0.12}_{-0.09}$ &   &  &  &  &  & 143.0  \\
                   & 20-Jun-1998  & 0.80/143 & $1.72^{+0.04}_{-0.05}$ & $ 1.46^{+ 0.38}_{- 0.18}$ & $2.08^{+0.03}_{-0.03}$ &   &  &  &  &  &  49.2  \\
                   & 29-Jun-1998  & 1.00/145 & $1.65^{+0.03}_{-0.03}$ & $ 2.71^{+ 0.46}_{- 0.30}$ & $2.05^{+0.07}_{-0.04}$ &   &  &  &  &  & 103.0  \\
                   & 16-Jul-1998  & 1.10/146 & $1.69^{+0.06}_{-0.05}$ & $ 1.52^{+ 0.73}_{- 0.46}$ & $2.23^{+0.09}_{-0.06}$ &   &  &  &  &  &  90.5  \\
                   & 25-Jul-1998  & 1.02/146 & $1.73^{+0.03}_{-0.03}$ & $ 1.59^{+ 0.41}_{- 0.29}$ & $2.17^{+0.04}_{-0.03}$ &   &  &  &  &  &  92.8  \\
                   & 10-Jun-1999  & 1.09/146 & $2.00^{+0.02}_{-0.02}$ & $ 1.12^{+ 0.06}_{- 0.06}$ & $2.45^{+0.02}_{-0.02}$ &   &  &  &  &  &  30.2  \\
H1722+119          & 24-Aug-2001  & & & & & 1.19/145 & $ 11.8^{+ 1.3}_{- 1.3}$ & $2.52^{+0.03}_{-0.08}$ & $ 6.90^{+ 0.74}_{- 2.72}$ & $3.21^{+0.53}_{-0.65}$ &   35.2  \\
1ES1959+650        & 04-May-1997  & 1.02/45  & $1.71^{+0.42}_{-1.12}$ & $ 1.23^{+ 0.68}_{- 0.37}$ & $2.67^{+0.11}_{-0.11}$ &   &  &  &  &  &  13.8  \\
                   & 23-Sep-2001  & 1.35/146 & $1.94^{+0.11}_{-0.13}$ & $ 0.33^{+ 0.31}_{- 0.33}$ & $2.50^{+0.05}_{-0.05}$ &   &  &  &  &  &  83.4  \\
                   & 28-Sep-2001  & 1.44/143 & $1.35^{+0.10}_{-0.12}$ & $ 1.11^{+ 0.06}_{- 0.05}$ & $2.31^{+0.02}_{-0.02}$ &   &  &  &  &  & 106.0  \\        
PKS2005-489        & 01-Nov-1998  & 1.74/146 & $2.13^{+0.02}_{-0.02}$ & $ 2.59^{+ 2.29}_{- 0.88}$ & $2.22^{+0.02}_{-0.02}$ &   &  &  &  &  & 177.0  \\
PKS2155-304        & 20-Nov-1996  & 0.95/228 & $2.36^{+0.03}_{-0.02}$ & $ 1.55^{+ 0.27}_{- 0.20}$ & $2.74^{+0.03}_{-0.02}$ &   &  &  &  &  &  55.0  \\
                   & 22-Nov-1997  & 1.40/145 & $2.28^{+0.01}_{-0.02}$ & $ 1.77^{+ 0.22}_{- 0.19}$ & $2.91^{+0.04}_{-0.03}$ &   &  &  &  &  &  83.3  \\
                   & 04-Nov-1999  & 1.07/145 & $2.59^{+0.03}_{-0.01}$ & $ 2.22^{+ 0.72}_{- 0.35}$ & $3.00^{+0.09}_{-0.04}$ &   &  &  &  &  &  24.7  \\
1ES2344+514        & 03-Dec-1999  & 1.48/91  & $1.72^{+0.19}_{-0.78}$ & $ 1.58^{+ 1.05}_{- 0.60}$ & $2.30^{+0.05}_{-0.05}$ &   &  &  &  &  &   8.9  \\
H2356-309          & 21-Jun-1998  & 1.21/145 & $1.56^{+0.11}_{-0.07}$ & $ 1.04^{+ 0.30}_{- 0.12}$ & $2.11^{+0.04}_{-0.04}$ &   &  &  &  &  &  24.3  \\
\hline																	  	     
\end{tabular}
}																
\end{center}																
\end{table*}

\begin{table*}																
\begin{center}
\rotate{
\scriptsize
\begin{tabular}{lccccccccccc}
\hline
\noalign{\smallskip}
\hline
\noalign{\smallskip}
&&\multicolumn{4}{c}{$N_{\rm H}$ FIXED }&\multicolumn{5}{c}{$N_{\rm H}$ FREE}&\\ 
\cline{4-7} \cline{8-12}\\
~~~~~Obj. Name     & Obs. Date   & $\chi^{2}_{r}$/d.o.f. & $\Gamma_{1}$   & $E_{break}$   & $\Gamma_{2}$  & $\chi^{2}_{r}$/d.o.f.& $N_{\rm H}$ &$\Gamma_{1}$   & $E_{break}$   & $\Gamma_{2}$  & Flux  \\
~~~~~~~~~~(1)      & (2)       &  (3)   &  (4)      & (5)      &   (6)       &  (7)      & (8)     & (9)    & (10)    & (11)   & (12)    \\
\noalign{\smallskip}
\hline
\noalign{\smallskip}																	     
\multicolumn{12}{c}{LBLs}\\																     
\noalign{\smallskip}
\hline
\noalign{\smallskip}
3C66A              & 31-Jan-1999  & 1.31/33  & $2.20^{+0.12}_{-0.27}$ & $ 0.26^{+ 1.54}_{- 1.07}$ & $2.32^{+0.15}_{-0.15}$ &   &  &  &  &  &   2.2  \\
S50716+71          & 14-Nov-1996  & 1.07/48  & $2.70^{+0.13}_{-0.14}$ & $ 2.35^{+ 0.80}_{- 1.23}$ & $1.96^{+0.16}_{-0.22}$ &   &  &  &  &  &   1.5  \\
                   & 07-Nov-1998  & 1.92/33  & $2.36^{+0.22}_{-0.37}$ & $ 1.75^{+ 3.99}_{- 1.75}$ & $1.84^{+0.26}_{-0.85}$ &   &  &  &  &  &   2.7  \\
                   & 30-Oct-2000  & 0.98/33  & $2.82^{+0.11}_{-0.12}$ & $ 2.73^{+ 0.42}_{- 1.49}$ & $1.78^{+0.23}_{-0.30}$ &   &  &  &  &  &   3.3  \\
ON231              & 11-May-1998  & 1.19/33  & $2.58^{+0.06}_{-0.06}$ & $ 3.09^{+ 1.03}_{- 1.38}$ & $1.58^{+0.28}_{-0.44}$ &   &  &  &  &  &   4.2  \\
                   & 11-Jun-1998  & 0.87/31  & $2.58^{+0.23}_{-0.09}$ & $ 2.70^{+ 0.47}_{- 1.33}$ & $1.49^{+0.19}_{-0.22}$ &   &  &  &  &  &   3.4  \\
BLLAC              & 31-Oct-2000  & 1.29/146 & $2.21^{+0.12}_{-0.13}$ & $ 0.45^{+ 0.39}_{- 0.44}$ & $2.64^{+0.05}_{-0.05}$ &   &  &  &  &  &  20.2  \\
\hline
\noalign{\smallskip}
\multicolumn{12}{c}{FSRQs}\\
\noalign{\smallskip}
\hline
\noalign{\smallskip}
1ES0836+710        & 27-May-1998  & 1.14/146 & $1.08^{+0.15}_{-0.23}$ & $ 0.38^{+ 0.37}_{- 0.37}$ & $1.34^{+0.04}_{-0.02}$ &   &  &  &  &  &  26.3  \\
3C273              & 18-Jul-1996  & 1.10/229 & $2.12^{+0.83}_{-1.19}$ & $ 0.37^{+ 0.38}_{- 0.07}$ & $1.58^{+0.01}_{-0.01}$ &   &  &  &  &  &  70.1  \\
                   & 24-Jun-1998  & 1.22/146 & $1.97^{+0.09}_{-0.06}$ & $ 0.72^{+ 0.11}_{- 0.13}$ & $1.61^{+0.02}_{-0.01}$ &   &  &  &  &  & 108.0  \\
                   & 09-Jan-2000  & 1.13/146 & $1.90^{+0.07}_{-0.05}$ & $ 0.84^{+ 0.22}_{- 0.16}$ & $1.63^{+0.01}_{-0.01}$ &   &  &  &  &  & 116.0  \\
                   & 13-Jun-2000  & 0.98/146 & $2.07^{+0.23}_{-0.12}$ & $ 0.58^{+ 0.20}_{- 0.15}$ & $1.64^{+0.02}_{-0.01}$ &   &  &  &  &  &  86.7  \\
                   & 12-Jun-2001  & 1.06/146 & $2.21^{+0.08}_{-0.07}$ & $ 0.82^{+ 0.11}_{- 0.06}$ & $1.65^{+0.02}_{-0.02}$ &   &  &  &  &  & 101.0  \\
PKS1510-089        & 03-Aug-1998  & 0.80/32  & $2.51^{+0.30}_{-0.30}$ & $ 1.41^{+ 1.21}_{- 0.25}$ & $1.32^{+0.10}_{-0.10}$ &   &  &  &  &  &   5.5  \\
PKS2126-158        & 24-May-1999  & 0.98/146 & $1.03^{+0.12}_{-0.12}$ & $ 0.60^{+ 6.66}_{- 0.58}$ & $1.78^{+0.27}_{-0.11}$ &   &  &  &  &  &  10.7  \\
\hline
\multicolumn{12}{l}{{\bf Columns}: {\bf 1}=Object Name; {\bf 2}=Observation Date; {\bf 3}=
Reduced $\chi^{2}$ and degrees of freedom; {\bf 4}= First photon spectral index; {\bf 5}=Energy break }\\
\multicolumn{12}{l}{in keV; {\bf 6}=Second photon spectral index. Columns 3 to 6 refer to fits 
obtained using the broken power law or the continuously curved }\\
\multicolumn{12}{l}{parabola model with intrinsic absorption fixed to the Galactic value; {\bf 7}=Reduced 
$\chi^{2}$ and degrees of freedom; {\bf 8}=Intrinsic absorption; {\bf 9}=First  }\\
\multicolumn{12}{l}{photon spectral index; {\bf 10}=Energy break in keV; {\bf 11}=Second 
photon spectral index. Columns 7 to 11 refer to fits obtained using the broken }\\
\multicolumn{12}{l}{power law or the continuously curved parabola model with intrinsic 
absorption free to vary. The intrinsic absorption is in units of $10^{20}{\rm cm}^{-2}$; }\\
\multicolumn{12}{l}{{\bf 12}=X--ray unabsorbed flux in the energy 
range 2--10 keV (in units of $10^{-12}$\flux).}\\
\end{tabular}
}
\end{center}
\end{table*}

\begin{table*}
\caption{Single power law fits for Spectra in Table~\ref{tab:bestfit2}} 
\label{tab:powerlaw}
\begin{center}
\scriptsize
\begin{tabular}{lcccc}
\hline
\noalign{\smallskip}
\hline
\noalign{\smallskip}
~~~~~Obj. Name     & Obs. Date   &$\chi^{2}_{r}$/d.o.f.& $N_{\rm H}$ &$\Gamma$    \\
~~~~~~~~~~(1)      & (2)          &  (3)     &  (4)                  &    (5)     \\
\noalign{\smallskip}
\hline
\noalign{\smallskip}
\multicolumn{5}{c}{HBLs}\\
\noalign{\smallskip}
\hline
\noalign{\smallskip}
PKS0548-322        & 20-Feb-1999  & 1.29/95  & $  3.2^{+  0.8}_{-  0.6}$ & $2.12^{+0.07}_{-0.06}$ \\
1ES1101-232        & 04-Jan-1997  & 0.97/257 & $  2.4^{+  3.2}_{-  2.0}$ & $2.01^{+0.06}_{-0.06}$ \\
                   & 19-Jun-1998  & 1.19/92  & $  3.5^{+  2.4}_{-  1.6}$ & $2.25^{+0.05}_{-0.05}$ \\
Mkn421             & 29-Apr-1997  & 2.28/239 & $  2.5^{+  2.5}_{-  2.5}$ & $2.65^{+0.02}_{-0.02}$ \\
                   & 30-Apr-1997  & 2.07/239 & $  2.6^{+  2.6}_{-  2.6}$ & $2.67^{+0.02}_{-0.02}$ \\
                   & 01-May-1997  & 2.38/239 & $  2.4^{+  2.4}_{-  2.4}$ & $2.63^{+0.02}_{-0.02}$ \\
                   & 02-May-1997  & 1.81/239 & $  2.7^{+  0.1}_{-  0.2}$ & $2.68^{+0.01}_{-0.04}$ \\
                   & 03-May-1997  & 1.62/239 & $  2.4^{+  0.2}_{-  0.1}$ & $2.74^{+0.03}_{-0.04}$ \\
                   & 04-May-1997  & 1.45/239 & $  2.6^{+  0.2}_{-  0.2}$ & $2.90^{+0.04}_{-0.05}$ \\
                   & 05-May-1997  & 1.79/239 & $  2.5^{+  0.1}_{-  0.2}$ & $2.78^{+0.04}_{-0.03}$ \\
                   & 07-May-1997  & 1.14/46  & $  1.1^{+  0.2}_{-  0.2}$ & $2.42^{+0.06}_{-0.05}$ \\
                   & 21-Apr-1998  & 7.82/155 & $  2.0^{+  0.1}_{-  0.1}$ & $2.51^{+0.01}_{-0.01}$ \\
                   & 23-Apr-1998  & 8.41/155 & $  1.9^{+  0.1}_{-  0.1}$ & $2.37^{+0.01}_{-0.01}$ \\
                   & 22-Jun-1998  & 4.32/155 & $  2.1^{+  0.2}_{-  0.2}$ & $2.43^{+0.01}_{-0.01}$ \\
                   & 04-May-1999  & 5.93/155 & $  2.2^{+  0.1}_{-  0.1}$ & $2.76^{+0.01}_{-0.01}$ \\
                   & 26-Apr-2000  & 2.53/155 & $  1.5^{+  0.1}_{-  0.1}$ & $2.02^{+0.01}_{-0.01}$ \\
                   & 28-Apr-2000  & 7.00/155 & $  1.4^{+  0.1}_{-  0.1}$ & $2.00^{+0.01}_{-0.01}$ \\
                   & 30-Apr-2000  & 3.95/155 & $  1.3^{+  0.1}_{-  0.1}$ & $2.01^{+0.01}_{-0.01}$ \\
                   & 09-May-2000  & 3.73/155 & $  1.3^{+  0.1}_{-  0.1}$ & $2.09^{+0.01}_{-0.01}$ \\
RXJ1117.1+2014     & 13-Dec-1999  & 1.36/85  & $  3.0^{+  0.4}_{-  0.4}$ & $2.86^{+0.08}_{-0.06}$ \\
1ES1218+304        & 12-Jul-1999  & 0.94/95  & $  2.9^{+  0.5}_{-  0.4}$ & $2.51^{+0.05}_{-0.05}$ \\
1RXSJ141756.8+25   & 13-Jul-2000  & 1.12/94  & $  2.8^{+  0.8}_{-  0.7}$ & $2.23^{+0.06}_{-0.07}$ \\
                   & 23-Jul-2000  & 1.16/39  & $  2.1^{+  1.3}_{-  1.2}$ & $2.22^{+0.15}_{-0.15}$ \\
1ES1517+656        & 05-Mar-1997  & 1.76/40  & $ 16.3^{+  5.8}_{-  4.2}$ & $2.49^{+0.13}_{-0.13}$ \\
1ES1553+113        & 05-Feb-1998  & 1.13/68  & $  6.9^{+  5.3}_{-  4.4}$ & $2.33^{+0.18}_{-0.17}$ \\
Mkn501             & 07-Apr-1997  & 1.15/236 & $  1.3^{+  0.1}_{-  0.1}$ & $1.90^{+0.01}_{-0.02}$ \\
                   & 11-Apr-1997  & 1.04/233 & $  0.9^{+  0.1}_{-  0.1}$ & $1.80^{+0.02}_{-0.02}$ \\
                   & 16-Apr-1997  & 1.41/236 & $  0.9^{+  0.1}_{-  0.1}$ & $1.59^{+0.01}_{-0.01}$ \\
                   & 28-Apr-1998  & 1.49/147 & $  1.0^{+  0.1}_{-  0.2}$ & $1.83^{+0.02}_{-0.02}$ \\
                   & 29-Apr-1998  & 1.37/142 & $  1.1^{+  0.1}_{-  0.1}$ & $1.82^{+0.02}_{-0.02}$ \\
                   & 01-May-1998  & 1.74/144 & $  1.4^{+  0.2}_{-  0.1}$ & $1.98^{+0.03}_{-0.02}$ \\
                   & 20-Jun-1998  & 1.07/147 & $  1.2^{+  0.2}_{-  0.2}$ & $2.03^{+0.03}_{-0.03}$ \\
                   & 29-Jun-1998  & 1.25/147 & $  1.4^{+  0.1}_{-  0.2}$ & $1.96^{+0.03}_{-0.03}$ \\
                   & 16-Jul-1998  & 1.69/147 & $  1.9^{+  0.3}_{-  0.2}$ & $2.07^{+0.03}_{-0.03}$ \\
                   & 25-Jul-1998  & 1.93/147 & $  1.5^{+  0.1}_{-  0.2}$ & $2.06^{+0.02}_{-0.03}$ \\
                   & 10-Jun-1999  & 1.78/147 & $  1.3^{+  0.1}_{-  0.1}$ & $2.41^{+0.01}_{-0.02}$ \\
H1722+119          & 24-Aug-2001  & 1.29/147 & $ 12.6^{+  1.3}_{-  1.3}$ & $2.56^{+0.02}_{-0.03}$ \\
1ES1959+650        & 04-May-1997  & 1.02/45  & $ 13.3^{+  7.5}_{-  6.2}$ & $2.70^{+0.11}_{-0.12}$ \\
                   & 23-Sep-2001  & 1.35/147 & $ 12.5^{+  2.1}_{-  2.0}$ & $2.51^{+0.05}_{-0.05}$ \\
                   & 28-Sep-2001  & 1.28/144 & $ 20.9^{+  0.9}_{-  1.0}$ & $2.33^{+0.02}_{-0.01}$ \\
PKS2005-489        & 01-Nov-1998  & 1.95/145 &                           & $2.19^{+0.01}_{-0.01}$ \\
PKS2155-304        & 20-Nov-1996  & 2.18/231 & $  1.3^{+  0.1}_{-  0.1}$ & $2.63^{+0.01}_{-0.01}$ \\
                   & 22-Nov-1997  & 5.70/147 & $  2.0^{+  0.1}_{-  0.1}$ & $2.70^{+0.01}_{-0.01}$ \\
                   & 04-Nov-1999  & 2.22/147 & $  0.9^{+  0.2}_{-  0.2}$ & $2.80^{+0.01}_{-0.01}$ \\
1ES2344+514        & 03-Dec-1999  & 1.41/92  & $ 11.8^{+  4.1}_{-  3.6}$ & $2.33^{+0.05}_{-0.05}$ \\
H2356-309          & 21-Jun-1998  & 1.33/147 & $  2.1^{+  0.4}_{-  0.4}$ & $2.09^{+0.03}_{-0.04}$ \\
\noalign{\smallskip}
\hline
\noalign{\smallskip}
\multicolumn{5}{c}{LBLs}\\
\noalign{\smallskip}
\hline
\noalign{\smallskip}
3C66A              & 31-Jan-1999  & 1.26/35  &                           & $2.26^{+0.18}_{-0.17}$ \\
S50716+71          & 14-Nov-1996  & 1.18/49  & $  0.1^{+  0.1}_{-  0.1}$ & $2.06^{+0.14}_{-0.13}$ \\
                   & 07-Nov-1998  & 2.06/35  &                           & $2.21^{+0.12}_{-0.14}$ \\
                   & 30-Oct-2000  & 1.55/34  &                           & $2.05^{+0.13}_{-0.12}$ \\
ON231              & 11-May-1998  & 2.86/35  &                           & $2.50^{+0.12}_{-0.14}$ \\
                   & 11-Jun-1998  & 3.04/34  &                           & $2.36^{+0.12}_{-0.14}$ \\
BLLAC              & 31-Oct-2000  & 1.28/147 & $ 10.1^{+  2.3}_{-  2.3}$ & $2.65^{+0.06}_{-0.05}$ \\ 
\noalign{\smallskip}
\hline
\end{tabular}
\end{center}
\end{table*}

\begin{table*}
\begin{center}
\scriptsize
\begin{tabular}{lcccc}
\hline
\noalign{\smallskip}
\hline
\noalign{\smallskip}
~~~~~Obj. Name     & Obs. Date   &$\chi^{2}_{r}$/d.o.f.& $N_{\rm H}$ &$\Gamma$    \\
~~~~~~~~~~(1)      & (2)          &  (3)     &  (4)                  &    (5)     \\
\noalign{\smallskip}
\hline
\noalign{\smallskip}
\multicolumn{5}{c}{FSRQs}\\
\noalign{\smallskip}
\hline
\noalign{\smallskip}
1ES0836+710        & 27-May-1998  & 1.13/147 & $ 78.0^{+ 54.9}_{- 35.2}$ & $1.34^{+0.04}_{-0.04}$ \\
3C273              & 18-Jul-1996  & 1.14/231 &                           & $1.58^{+0.01}_{-0.01}$ \\
                   & 24-Jun-1998  & 2.22/148 &                           & $1.65^{+0.01}_{-0.01}$ \\
                   & 09-Jan-2000  & 2.10/148 &                           & $1.67^{+0.01}_{-0.01}$ \\
                   & 13-Jun-2000  & 1.79/148 &                           & $1.68^{+0.01}_{-0.01}$ \\
                   & 12-Jun-2001  & 3.08/148 &                           & $1.74^{+0.01}_{-0.01}$ \\
PKS1510-089        & 03-Aug-1998  & 1.37/35  &                           & $1.37^{+0.09}_{-0.10}$ \\
PKS2126-158        & 24-May-1999  & 1.06/147 & $294.1^{+110.7}_{- 86.5}$ & $1.62^{+0.04}_{-0.04}$ \\ 
\noalign{\smallskip}
\hline
\end{tabular}
\end{center}
{\bf Columns}: {\bf 1}=Object Name; {\bf 2}=Observation Date; {\bf 3}=Reduced 
$\chi^{2}$ and degrees of freedom; {\bf 4}=Free intrinsic
absorption (in units of $10^{20}{\rm cm}^{-2}$); {\bf 5}=Photon spectral 
index for the single power law model. 
\end{table*}

\end{document}